\documentclass[12pt,a4paper]{iopart}
\usepackage{graphicx,color}
\usepackage{dcolumn}
\usepackage{epsfig}
\usepackage{setstack}

\usepackage{iopams}
\usepackage{euscript}
\usepackage{amssymb}
\usepackage{amsthm}
\usepackage{newlfont}
\usepackage{mathrsfs}
\usepackage{subfigure}
\usepackage{changes}

\newcommand{\opunit}{\text{1}\kern-0.22em\text{l}}

\newcommand{\ie}{\textit{i.e.}}

\newcommand{\id}{\rmd}

\def\bea{\begin{eqnarray}}
\def\eea{\end{eqnarray}}
\def\ba{\begin{array}}
\def\ea{\end{array}}
\def\n{\nonumber}
\def\la{\langle}
\def\ra{\rangle}

\def\ve{\varepsilon}

\def\Jd{J_\rmd}

\begin{document}

\title[Zero-current Nonequilibrium State]{Zero-current Nonequilibrium State  in Symmetric Exclusion Process with Dichotomous Stochastic Resetting}

\author{Onkar Sadekar}
\address{Indian Institute of Science Education and Research, 
\\ Homi Bhabha Road, Pashan, Pune 411008, India
}

\author{Urna Basu}
\address{Raman Research Institute, \\
C. V. Raman Avenue, Bengaluru 560080, India}

\begin{abstract}
 
 We study the dynamics of symmetric exclusion process (SEP) in the presence of stochastic resetting to two possible specific configurations ---
 with rate $r_1$ (respectively, $r_2$) the system is reset to a step-like configuration where all the particles are clustered in the left (respectively, right) half of the system. We show that this dichotomous resetting leads to a range of rich behaviour, both dynamical and in the stationary state. We calculate the exact stationary profile in the presence of this dichotomous resetting and show that the diffusive current grows linearly in time, but unlike the resetting to a single configuration, the current can have negative average value in this case.  For $r_1=r_2,$ the average current vanishes, and density profile becomes flat in the stationary state, similar to the equilibrium SEP. However, the system remains far from equilibrium and we characterize the nonequilibrium signatures of this `zero-current state'. We show that both the spatial and temporal density correlations in this zero-current state are radically different than in equilibrium SEP.
 We also study the behaviour of this zero-current state under an external perturbation and demonstrate that its response differs drastically from that of equilibrium SEP - while a small driving field generates a current which grows as $\sqrt{t}$ in the absence of resetting, the zero-current state in the presence of dichotomous resetting shows a current $\sim t$ under the same perturbation.  
\end{abstract}

\maketitle

\section{Introduction}\label{sec:intro}

Introduction of stochastic resetting changes the statistical properties of a system drastically. Recent years have seen a tremendous surge in studying the effect of resetting on a wide variety of systems \cite{EvansReview2019}. The paradigmatic example is that of a Brownian particle whose position is stochastically reset to a fixed point in space \cite{Brownian, Brownian2, Experiment_reset}. This results in a set of intriguing behavior like non-trivial stationary distribution, unusual dynamical relaxation behaviour and a finite mean first-passage time. Various generalizations and extensions of this model have been studied \cite{Brownian3,absorption,highd,Mendez2016,Puigdellosas,Arnab2019_2}; resetting in the presence of an external potential or confinement \cite{Arnab2015, potential, circle, interval}, resetting of an underdamped particle~\cite{deepak} and resetting to already excursed positions \cite{Boyer2014, Sanjib2015, Mendez2019} are some notable examples. 
Moreover, instead of a constant resetting rate, other resetting protocols have also been studied which lead to a wide range of novel statistical properties. Examples include space \cite{Roldan2017} and time-dependent resetting rates \cite{time-dep1,time-dep2}, non-Markovian resetting  \cite{ShamikPRE, nonmarkov2,nonmarkov3}, resetting followed by a refractory period  \cite{refractory,refractory2} and resetting with space-time coupled return protocols \cite{Reuveni2019}.

An important issue which has gained a lot of attention recently is the 
effect of resetting on extended systems with many interacting degrees of freedom. This question has been studied in the context of fluctuating surfaces~\cite{kpz-reset}, coagulation-diffusion processes~\cite{Durang2014}, symmetric exclusion process \cite{sep-reset}, zero-range process and its variants~\cite{Grange, Grange2} and Ising model \cite{ising-reset}. 
It has been shown that introduction of resetting leads to a wide range of novel phenomena in these systems. A particularly interesting question is how the presence of resetting affects the behaviour of current, which plays an important role in characterizing nonequilibrium stationary states of such extended systems. The exclusion processes \cite{Liggett, Spohn-book}, which refer to a class of simple and well studied models of interacting particles, are particularly suitable candidates for exploring these issues. It has been shown recently that for the simple symmetric exclusion process, the presence of stochastic resetting drastically affects the behaviour of the current \cite{sep-reset}. A natural question is what happens to the current  if more complex resetting protocols are used. A first step is to study a scenario with more than one resetting rates which take the system to different configurations. 


In this article we address this question with a dichotomous resetting protocol in the context of Symmetric Exclusion Process (SEP). 
We study the behaviour of SEP in the presence of stochastic resetting to either of two specific configurations with different rates $r_1$ and $r_2.$ The two resetting configurations are chosen to be two complementary step-like configurations where all the particles are concentrated in the left-half, respectively, right-half, of the system. We calculate the exact time-dependent and stationary density profile in the presence of this dichotomous resetting. We also investigate the behaviour of the diffusive current and show that, similar to the case with resetting to one configuration, in the long-time regime the current grows linearly with time. However, depending on whether $r_1$ is larger than $r_2$ or not, the average current can be positive or negative. Moreover, it turns out that, in the short-time regime, a superlinear temporal growth $\sim t^{3/2}$  of the average current can be observed depending on the choice of the initial condition. We also calculate the variance, skewness and kurtosis of the current distribution for small values of the resetting rates $r_1,r_2 \ll 1,$ and show that, in the long time limit, the fluctuations of the diffusive current are characterized by a Gaussian distribution.

For the special case $r_1=r_2,$ the stationary profile becomes flat and the diffusive particle current vanishes. We explore the question - how is this zero-current state (ZCS) different than equilibrium SEP? To answer this question, we explore the nature of the spatial and temporal correlations in the ZCS. It turns out that, the presence of resetting has a strong effect on both the spatial and temporal correlations of the system. We show that, even though flat, the density profile is strongly correlated, contrary to the equilibrium case.  On the other hand, the temporal auto-correlation in ZCS decays exponentially with time, as opposed to an algebraic decay in equilibrium. Moreover, we study the effect of an external perturbation on this state by adding a external drive along the central bond which biases the hopping rate across the bond. We show that the response of the ZCS is drastically different than that of ordinary SEP - while the current generated due to the perturbation grows as $\sim \sqrt{t}$ for equilibrium SEP, in the presence of the dichotomous resetting, the same perturbation leads to a linear temporal growth in the diffusive current.
 
 The structure of the paper is as follows. In the next Section we define the dynamics of simple exclusion process with dichotomous resetting and derive the corresponding renewal equation. The dynamical and stationary behaviour of the density profile is investigated in  Sec.~\ref{sec:density}. In  Sec.~\ref{sec:current} we study the behaviour of the diffusive current including its moments and distribution. Sec.~\ref{sec:zero_cur} is devoted to the study of the special scenario $r_1=r_2$:  How current fluctuations and configurations weights in ZCS differ from SEP is discussed in Sec.~\ref{sec:zcs_current} and \ref{sec:zcs_config}. In Sec.~\ref{subsec:spat_correl} and~\ref{subsec:time_correl} we explore the behaviour of the spatial and temporal correlation of the density. The response of the system to an external perturbation is investigated in Sec.~\ref{subsec:perturb}. We conclude with a summary of our results  in Sec.~\ref{sec:conclusions}.

\section{Model}\label{sec:model}

Let us consider a periodic lattice of size $L$ where each site $x$ can either be occupied by one particle or be vacant; correspondingly, the site variable $s_x=1,0$. Consequently, the configuration $\cal C$ of the system is characterized by an array $\cal C = \{s_x; x=0,1,\cdots, L-1 \}.$ Moreover, we consider the case of half-filling, \ie, the number of particles $\sum_x s_x = L/2.$ The configuration evolves following two different kinds of dynamical moves, namely, hopping and resetting. The hopping dynamics is the usual one for ordinary symmetric exclusion process -- a randomly chosen particle hops to one of its neighbouring sites with unit rate, provided the target site is empty. The stochastic resetting dynamics refers to an abrupt change in the configuration --  at any time, the system is `reset' to either of the two specific configurations $\cal C_1$ and $\cal C_2,$ with rates $r_1$ and $r_2,$ respectively. 
In this work we choose $\cal C_1$ and $\cal C_2$ to be two step-like configurations where all the particles are clustered in the left, respectively right, half of the system, \ie,
\bea
\fl \cal C_1 = \left \{ \begin{array}{cc}
                     s_x = 1 &\quad  0 \le x \le \frac L2 -1 \cr
                     s_x = 0 & \quad \textrm {otherwise}
                    \end{array}
                  \right., \qquad \textrm{and} \qquad
 \cal C_2 = \left \{ \begin{array}{cc}
                     s_x = 0 &\quad  0 \le x \le \frac L2 -1 \cr
                     s_x = 1 & \quad \textrm{otherwise}
                    \end{array}
                  \right..  \label{eq:C1C2}               
\eea
This stochastic resetting to either of the two complimentary configurations with different rates is referred to as `dichotomous resetting' in the rest of the paper. The behaviour of SEP with resetting to only one possible configuration was studied in Ref.~\cite{sep-reset}. In this work we will show that in the presence of dichotomous resetting the system shows a much more rich behaviour including an interesting nonequilibrium zero-current stationary state, whose properties we will characterize.

Let $\cal P(\cal C,t | \cal C_0,0)$ denote the probability that the configuration of the system is $\cal C$ at time $t,$ starting from some initial configuration $\cal C_0$ at time $t=0.$ The master equation governing the evolution of $\cal P(\cal C,t | \cal C_0,0)$ reads,
\bea
\fl \qquad  \quad\frac{\id}{\id t} \cal P (\cal C,t | \cal C_0,0)= \cal L_0 \cal P (\cal C,t | \cal C_0,0) - (r_1 +r_2)\cal P (\cal C,t| \cal C_0,0) +r_1 \delta_{\cal C,\cal C_1} + r_2\delta_{\cal C,\cal C_2}, \label{eq:ME_PC}
\eea
where $\cal L_0$ denotes the Markov Matrix in the absence of resetting. In other words, $\cal L_0 \cal P (\cal C,t) = \sum_\cal {C'} [ W_{\cal C' \rightarrow \cal C}\cal P(\cal {C'},t) - W_{\cal C \rightarrow \cal {C'}}\cal P(\cal C,t)]$ where $W_{\cal {C'} \rightarrow C}$ represents the jump rate from configuration $\cal {C'}$ to $\cal C$ due to hopping dynamics only. This rate is equal to $1$ only if $\cal C$ can be obtained from $\cal {C'}$ by a single hop of a particle to either of its vacant neighbouring sites. Equation~\eref{eq:ME_PC} can be formally solved to get,
\bea
\fl\qquad \cal P (\cal C,t| \cal C_0,0)=e^{(\cal L_0 -r)t}\cal \delta_{\cal C, \cal C_0} +r_1 \int_0^t ds \enskip e^{(\cal L_0 -r)s} \delta_{\cal C,\cal C_1} + r_2\int_0^t ds\enskip e^{(\cal L_0 -r)s} \delta_{\cal C,\cal C_2}, \label{eq:PC_sol1}
\eea
where $r=r_1+r_2$ denotes the sum of the two resetting rates. Now, let us note that $e^{\cal L_0 t} \delta_{\cal C,\cal C_i}$ is nothing but $\cal P_0(\cal C,t| \cal C_i,0)$ -- probability that the configuration is $\cal C$ at time $t,$ starting from some configuration $\cal C_i,$ {\it in the absence of resetting}. Equation \eref{eq:PC_sol1} then can be rewritten as,
\bea
\fl ~~\cal P (\cal C,t|\cal C_0,0)&=& e^{-rt}\cal P_0 (\cal C,t| \cal C_0,0) +r_1 \int_0^t \id s ~ e^{-rs} \cal P_0 (\cal C,s| \cal C_1,0) + r_2\int_0^t \id s~ e^{-rs} \cal P_0 (\cal C,s| \cal C_2,0) \cr
&=&e^{-rt}\cal P_0 (\cal C,t| \cal C_0,0) +r \int_0^t \id s ~ e^{-rs} [\alpha \cal P_0 (\cal C,s| \cal C_1,0) + (1-\alpha) \cal P_0 (\cal C,s| \cal C_2,0) ],\label{eq:PC_sol2}
\eea
where, in the last step, we have used $\alpha = \frac {r_1}{r_1+r_2}.$  

Equation \eref{eq:PC_sol2} is nothing but the renewal equation for the dichotomous resetting, which can be understood easily using  arguments similar to those used for single resetting configuration~\cite{Brownian,kpz-reset}. 
Let us consider the system at some time $t$ and let $s$ be the time elapsed since the last resetting event. Now, the probability that no resetting event occurred during this interval is given by $r e^{-rs}$ (let us recall that $r=r_1+r_2$) where $0 \le s \le t$ is a random variable. During the interval $[t-s,t]$ the system evolves following ordinary SEP dynamics, starting from $\cal C_1$, or $\cal C_2,$ depending on the last resetting configuration. Now, probability that the last resetting was to configuration $\cal C_1,$ respectively $\cal C_2,$ is nothing but $\alpha = \frac{r_1}r,$ respectively $1-\alpha = \frac{r_2}r.$ Then, the probability that the system is at configuration $\cal C$ at time $t$ with at least one resetting event occurring during $[0,t]$ is given by the integral in Eq.~\eref{eq:PC_sol2}. The first term corresponds to the scenario where no resetting event occurred during time $t$ (probability $e^{-rt}$) and the system evolved as ordinary SEP during the whole time.  

\begin{figure}[t]
 \centering
 \includegraphics[width=10 cm]{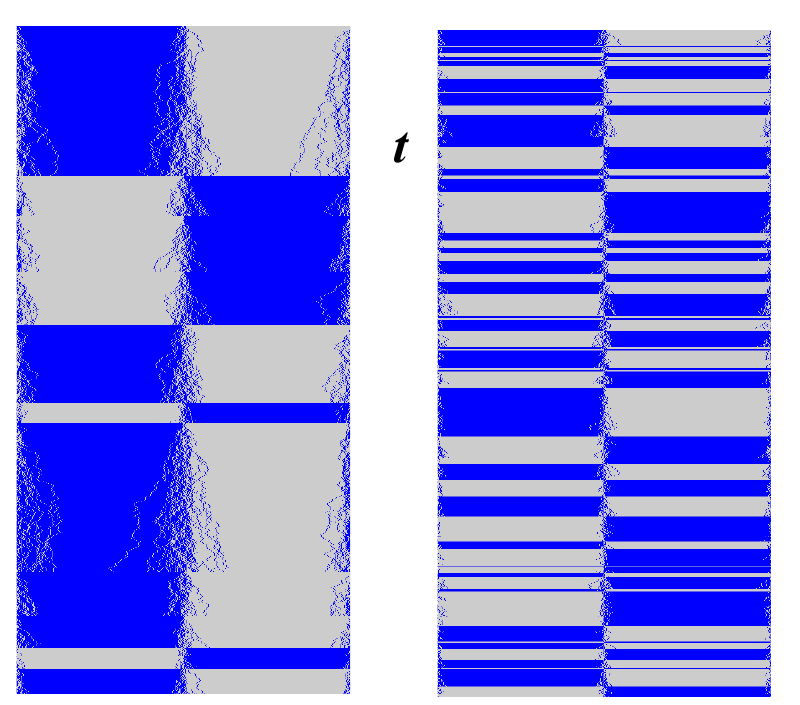}
 \caption{Typical snapshots for the SEP with dichotomous resetting for a system of size $L=400.$ The left panel shows the time-evolution for $r_1=r_2=0.005$ and the right panel shows the same for  $r_1=r_2=0.05.$ }
 \label{fig:snapshot}
\end{figure}

The typical time duration between two consecutive resetting events is $r^{-1},$ irrespective of which configuration it resets to. As mentioned above, during this time the system evolves following the hopping dynamics only. Figure \eref{fig:snapshot} shows typical time-evolution of the system for two different values of $r,$ with $r_1=r_2.$ In the following we study the behavior of SEP in the presence of dichotomous resetting.

\section{Density profile}\label{sec:density}

In this section, we investigate the dynamical and static behaviour of the density profile $\rho(x,t)= \la s_x(t) \ra$  in the presence of the dichotomous resetting. For resetting to only one configuration the exact time-evolution of the density profile has been calculated in Ref.~\cite{sep-reset}. Here we follow the same procedure to investigate the behaviour of the density profile for the dichotomous case.

The time evolution equation for $\rho(x,t)$  can be obtained by multiplying Eq.~\eref{eq:ME_PC} by $s_x$ and then summing over all possible configurations $\cal C$,
\bea
\fl \quad\frac{d}{dt}\rho (x,t) = \rho (x+1,t) + \rho (x-1,t) - 2\rho (x,t)  - r\rho(x,t) +r_1\phi_1(x) + r_2\phi_2(x). \label{eq:drdt}
\eea
Here, $\phi_1(x)$ and $\phi_2(x)$ are the density profiles corresponding to the resetting configurations $\cal C_1$ and $\cal C_2$ respectively. Equation \eref{eq:drdt} can be solved using the Fourier transform of $\rho (x,t)$,
\bea
\tilde\rho (n,t)= \displaystyle \sum_{x=0}^{L-1} e^{i \frac{2 \pi nx}{L}} \rho (x,t) \qquad {\text with} \enskip n=0,1,2,...,L-1. \label{eq:rho_FT}
\eea
Substituting Eq. \eref{eq:rho_FT} in Eq.~\eref{eq:drdt} we get,
\bea
\frac{d}{dt}\tilde\rho (n,t)= -(\lambda_n + r)\tilde\rho (n,t) +  r_1\tilde\phi_1 (n) + r_2\tilde\phi_2 (n),\label{eq:drndt}
\eea
where $\lambda_n= 2\left(1-\cos\frac{2{\pi}n}{L}\right)$ and $\tilde\phi_1 (n)$ and $\tilde\phi_2(n)$ are the Fourier transforms of $\phi_1(x)$ and $\phi_2(x)$ respectively. Eq. \eref{eq:drndt} can be solved immediately to obtain,
\bea
\tilde\rho (n,t)= \left[\tilde\phi_0 (n) - \frac{r_1\tilde\phi_1 (n) + r_2\tilde\phi_2 (n)}{\lambda_n + r}\right]e^{-(\lambda_n + r)t} + \frac{r_1 \tilde\phi_1 (n) + r_2 \tilde\phi_2 (n)}{\lambda_n + r}.\label{eq:rnt}
\eea
Here, $\tilde\phi_0 (n)$ denotes the Fourier transform of initial density profile $\rho(x,0)$. We can obtain $\rho(x,t)$ from Eq.~\eref{eq:rnt} by taking the inverse Fourier transform of $\tilde\rho (n,t),$ 
%
\bea
\fl \qquad \rho (x,t)=\frac{1}{L} \displaystyle \sum_{n=0}^{L-1}\left[\tilde\phi_0 (n) e^{-(\lambda_n + r)t}
+ \frac{r_1\tilde\phi_1 (n) + r_2\tilde\phi_2 (n)}{\lambda_n + r}\bigg( 1-e^{-(\lambda_n + r)t} \bigg) \right] e^{-i\frac{2 \pi  nx}{L}}.
\label{eq:rxt_sol}
\eea 
\begin{figure}[t]
 \centering
 \includegraphics[width=8 cm]{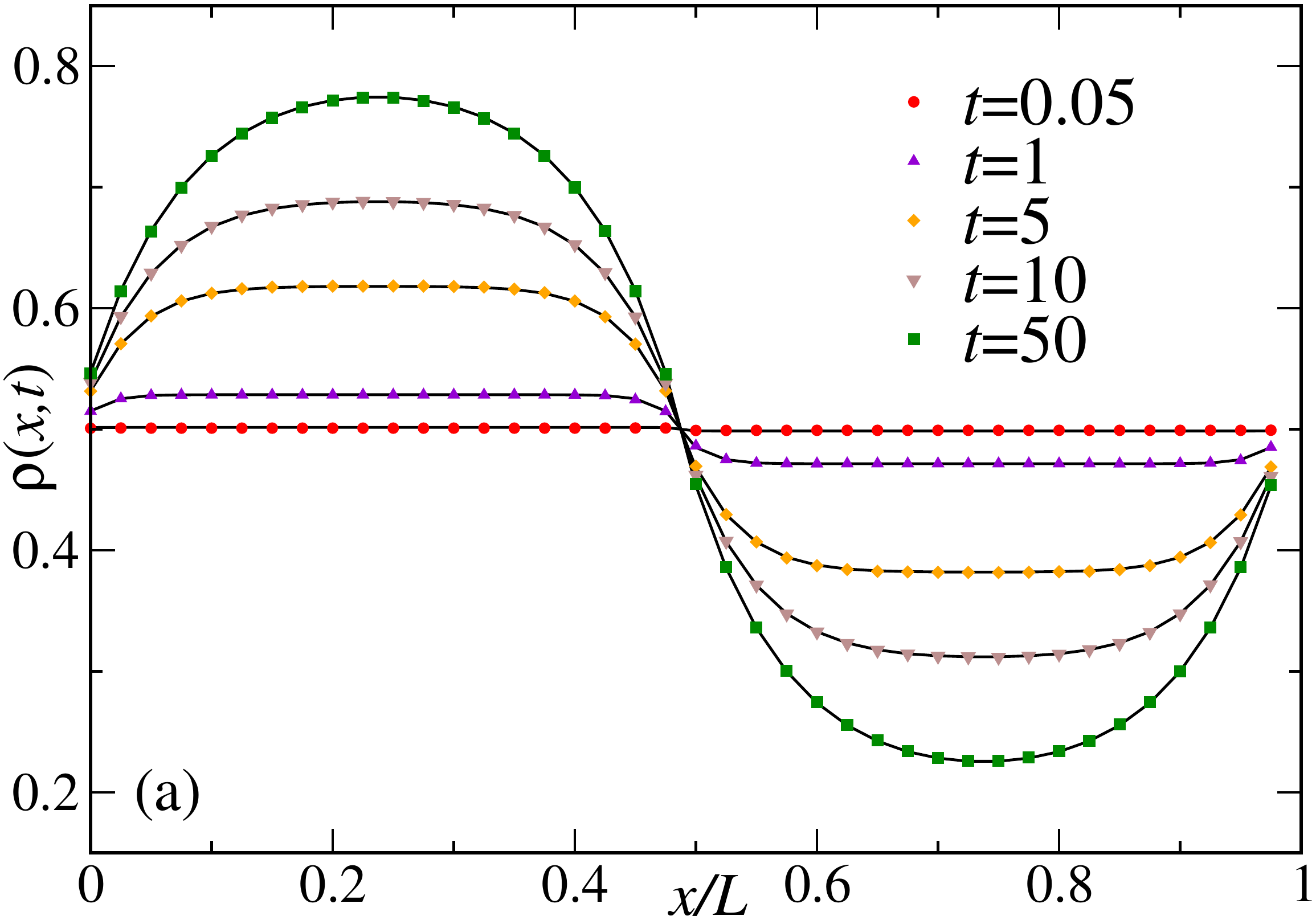}\includegraphics[width=8 cm]{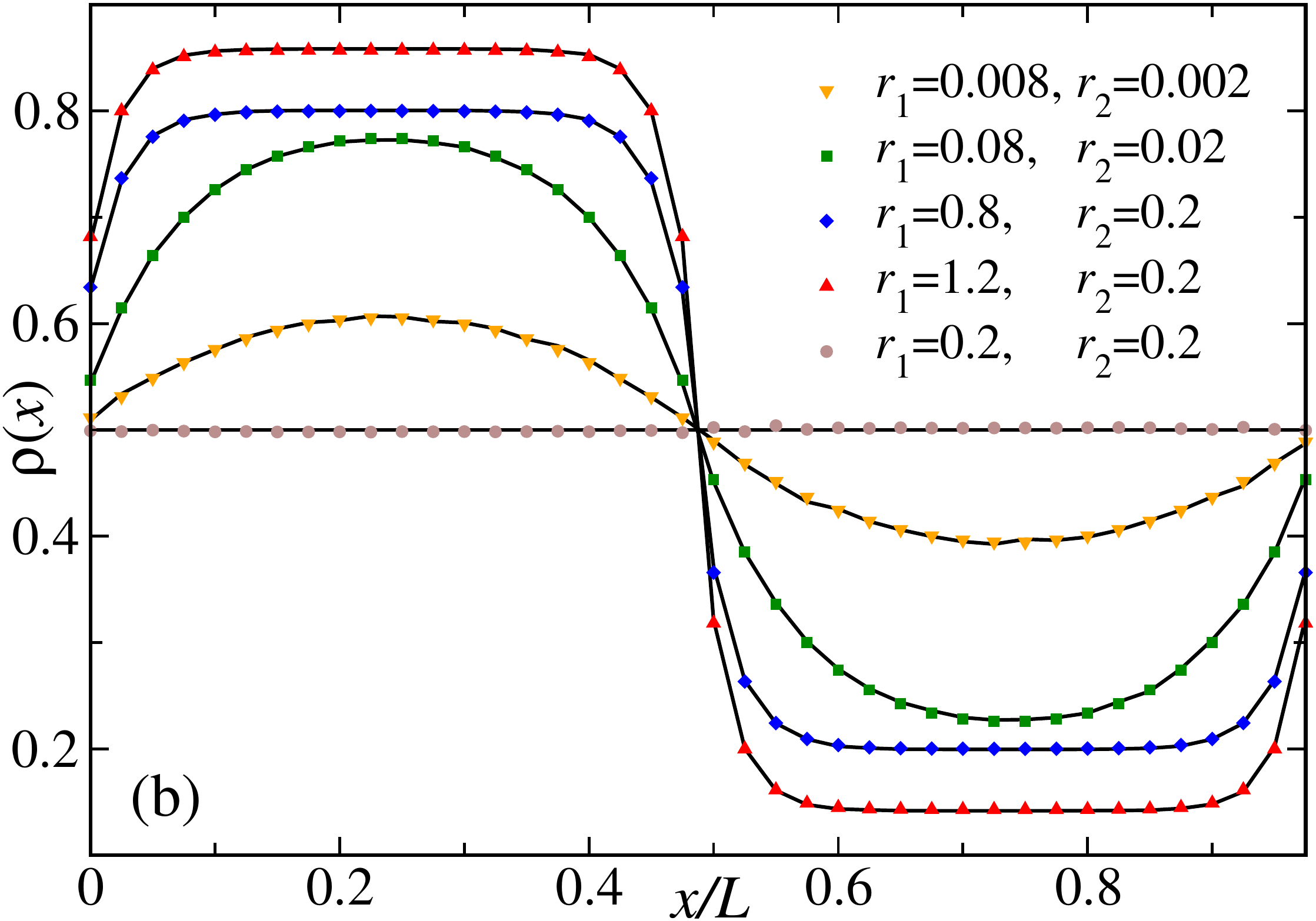}
 \caption{Density profile: (a) Evolution of $\rho(x,t)$ as a function of time $t$ for $r_1=0.08$ and $r_2=0.02.$  
 (b) Plot of the stationary profile $\rho(x)$ as a function of $x/L$ for different values of $r_1$ and $r_2.$  The solid lines correspond to the analytical results whereas the symbols correspond to the data obtained from numerical simulations. The system size $L=40$ here.}
 \label{fig:st_profile}
\end{figure}
It is to be noted that $\rho(x,t)$ also satisfies a renewal equation which can easily be derived from Eq.~\eref{eq:PC_sol2},
\bea
\rho(x,t)= e^{-rt} \rho^0_0(x,t) + r \int_0^t \id s~ e^{-r s} \bigg[\alpha \rho_0^1(x,t) + (1-\alpha)\rho_0^2(x,t) \bigg].\label{eq:rho_renewal}
\eea
Here $\rho_0^i(x,t)$ denotes the density at time $t,$ starting from the initial profile $\phi_i(x),$ for ordinary SEP (the subscript $0$ refers to the absence of resetting). 
Recalling that, $\rho_0^i(x,t) = \frac 1L \sum_{n=0}^{L-1} e^{-i \frac{2 \pi n x}{L}}e^{-\lambda_n t} \tilde \phi_i(n)$ (see Ref~\cite{sep-reset}), it is straightforward to show that Eq.~\eref{eq:rho_renewal} is equivalent to Eq.~\eref{eq:rxt_sol}.

To proceed further we need to specify the initial condition. Let us consider the case where initially the system starts from $\cal C_1$ and $\cal C_2$ with equal probability.
In that case, $\rho(x,0)=\phi_0(x)=\frac{1}{2}[\phi_1(x)+\phi_2(x)].$ For our specific choice of $\cal C_{1,2}$ (see Eq.~\eref{eq:C1C2}) we have $\phi_1(x)= \Theta(\frac L2-1-x)$ and $\phi_2(x)=\Theta(x-\frac L2)$ where the Heaviside Theta function is defined as $\Theta(x) = 1 $ for $x \ge 0$ and it is zero elsewhere. Consequently, $\phi_1(x)+\phi_2(x)=1~ \forall x$ and we have $\tilde \phi_0(n) = \frac L2 \delta_{n0}.$  Additionally, we also have, $\tilde\phi_i (0)=\frac{L}{2}, \, i=1,2$. For $n \neq 0, \tilde\phi_2(n)=-\tilde\phi_1(n)$. It is easy to see that (see Ref.~\cite{sep-reset}),
\bea
\tilde\phi_1(n) &=& \left\{ 
\begin{array}{cl}
1+i\cot \frac{{\pi}n}{L} & \quad \textrm{for}~ n=1,3,5,...\cr
0 &   \quad \textrm{for} \;\; \textrm{even}~n \ge 2.                          
\end{array}
\right. \label{eq:phi_tilde}
\eea
Substituting the above results in Eq. \eref{eq:rxt_sol} and simplifying we get,
\bea
\fl \;\; \rho (x,t) = \frac{1}{2} + \frac{1}{L}\left[ \displaystyle \sum_{n=1,3}^{L-1}  \left(\frac{r_1 - r_2 }{\lambda_n + r_1 + r_2}\right)\left(1+i\cot\frac{\pi n}{L}\right) e^{-i\frac{2{\pi}nx}{L}} (1-e^{-(\lambda_n + r)t})\right]. \label{eq:rxt_sol_sim}
\eea
Note that, for any value of $x$ and $t$,
$\rho (x,t ;r_1,r_2)=\frac{1}{2}-\rho (x,t; r_2,r_1)$. For this reason, it suffices to study only the regime $r_1 \ge r_2.$ Figure \ref{fig:st_profile}(a) shows the time-evolution of $\rho(x,t)$ for $r_1=0.08$ and $r_2=0.02$; starting from a flat profile at time $t=0,$ where the density at each site equals $\frac 12,$ the average density profile evolves to a non-trivial inhomogeneous  stationary state. The stationary profile can be obtained by taking the $t \to \infty$ limit in Eq.~\eref{eq:rxt_sol_sim} and yields,
\bea
\rho (x) = \frac{1}{2} + \frac{1}{L}\left[ \displaystyle \sum_{n=1,3}^{L-1}  \left(\frac{r_1 - r_2 }{\lambda_n + r}\right)\left(1+i\enskip \cot\frac{\pi n}{L}\right) e^{-i\frac{2{\pi}nx}{L}} \right]. \label{eq:rx_sol_st}
\eea
Figure \ref{fig:st_profile}(b) shows plots of $\rho(x)$ for a set of different values of $(r_1,r_2).$

A special situation arises when the two resetting rates are equal, \ie, $r_1=r_2=\frac{r}{2}.$ In this case, as can be seen from  Eq.~\eref{eq:rx_sol_st}, the stationary profile remains flat irrespective of the value of $r.$ \footnote{In fact, for our specific choice of initial condition, for $r_1=r_2,$ $\rho(x,t)= \frac 12$ at all times $t \ge 0.$} This is reminiscent of the ordinary SEP in equilibrium. However, in the presence of the dichotomous resetting, the system remains far away from equilibrium  although it is not apparent from the density profile alone. We will come back to this question later in Sec.~\ref{sec:zero_cur}.

\section{Diffusive Current}\label{sec:current}

\begin{figure}[t]
 \centering
 \includegraphics[width=9 cm]{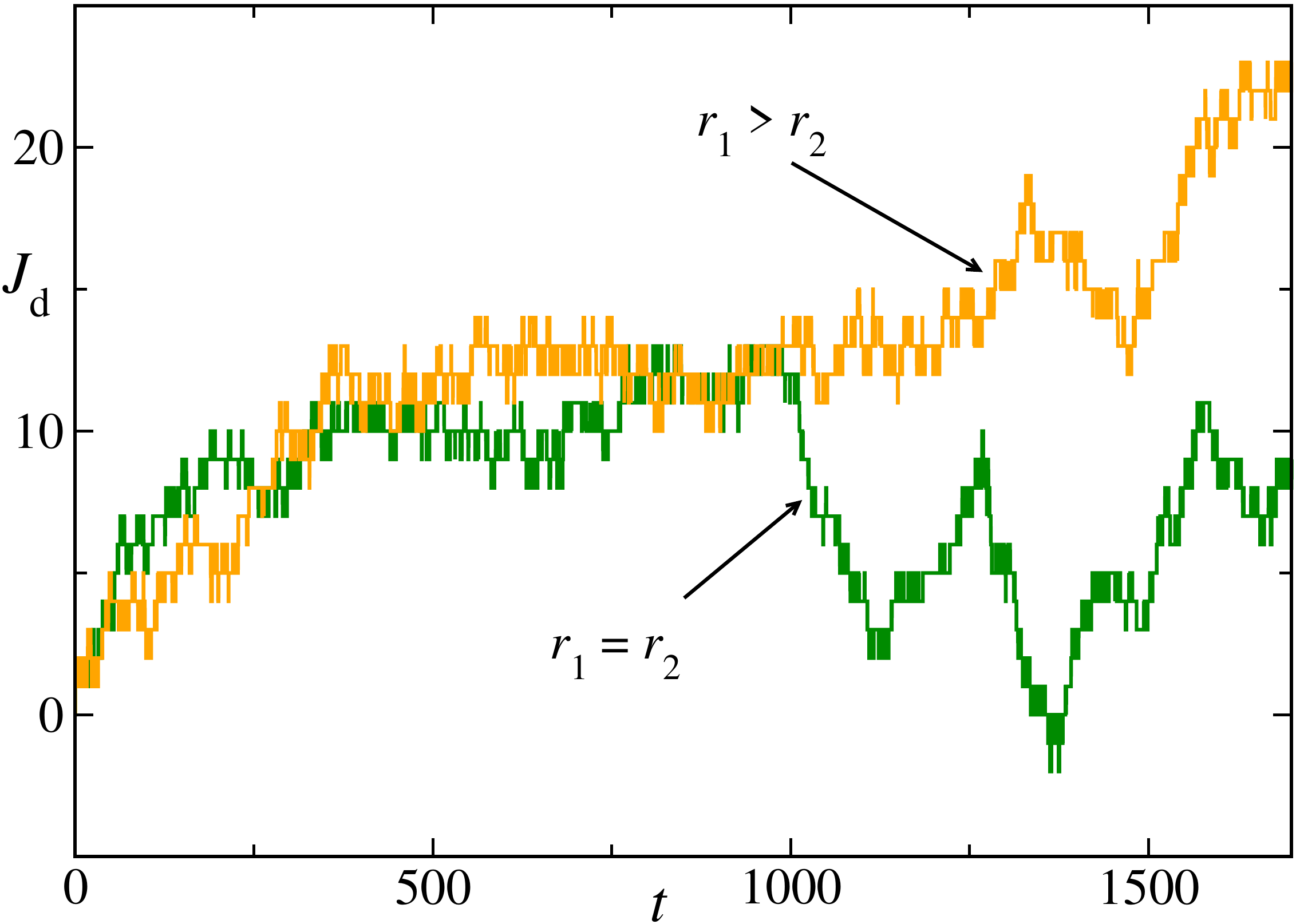}
 \caption{Two typical trajectories for $\Jd$ for $r_1=r_2$ and $r_1>r_2$ respectively. Note that the trajectory for $r_1<r_2$ would be qualitatively same as the $r_1>r_2$ case except for overall decrease instead of increase in $\Jd$.}
 \label{fig:traj_cur}
\end{figure}

In the absence of resetting, starting from either of the two step-like configurations $\cal C_1$ and $\cal C_2,$ the hopping of particles results in a diffusive particle current. Quantified by the net number of particles crossing the central bond up to time $t,$ this diffusive current increases $\sim \sqrt{t}$ for a thermodynamically large system~\cite{Derrida1,Derrida2}. It has been shown that the presence of stochastic resetting to only one configuration, namely $\cal C_1,$ alters the behaviour of the diffusive current drastically,  resulting in a linear growth in time \cite{sep-reset}. In that case, because of the choice of the resetting configuration, the net motion of the particles always remains from the left to the right-half of the system. The presence of the dichotomous resetting can change that, as resetting to $C_2$ would give rise to particles moving to the left-half from the right-half of the system. Figure \ref{fig:traj_cur} shows a plot of the diffusive current $\Jd$ along a couple of typical trajectories for $r_1$ larger than and equal to $r_2;$ the sudden increase (decrease) in $\Jd$ indicates a resetting to $\cal C_1$ ($\cal C_2$). In this section we quantitatively characterize the behaviour of the diffusive current in the presence of dichotomous resetting, for arbitrary values of $r_1,r_2.$

The instantaneous diffusive current $j_\id(t)$ at time $t$ is defined as the net number of particles crossing the central bond $\left(\frac{L}{2}-1,\frac{L}{2} \right)$ from left to right, in the time interval $(t,t+dt)$. The average instantaneous current, then, is nothing but the density gradient across the central bond,
\bea
\fl \qquad \la j_\id(t) \ra = \left\la s_{\frac{L}{2}-1}\left(1-s_{\frac{L}{2}}\right)\right\ra - \left\la s_{\frac{L}{2}}\left(1-s_{\frac{L}{2}-1}\right)\right\ra =\rho \left({\frac{L}{2}-1},t \right) - \rho \left({\frac{L}{2}},t\right).\label{eq:cur_def}
\eea
Using the expression for $\rho (x,t)$ from Eq. \eref{eq:rxt_sol_sim} in Eq. \eref{eq:cur_def} we get,
\bea
\la j_\id(t) \ra = \frac{2 (r_1 - r_2)}{L}\displaystyle \sum_{n=1,3}^{L-1}  \frac{1-e^{-(\lambda_n + r)t}}{\lambda_n + r}.  \label{eq:jdt}
\eea
Clearly, the average instantaneous current becomes negative when $r_1< r_2.$

We are interested in the behaviour of the total time-integrated current $\Jd(t)$ which measures the total number of particles crossing the central bond up to time $t.$ The average total current $\la \Jd(t) \ra$ can be obtained by integrating the average instantaneous current $\la J_d(t) \ra= \int_0^t \id t' \; \la j_\id(t') \ra.$ Integrating Eq. \eref{eq:jdt} we get, 
\bea
\la J_d(t) \ra= \frac{2 (r_1 - r_2)}{L} \displaystyle \sum_{n=1,3}^{L-1} \left[ \frac{t}{\lambda_n + r} -\frac{(1-e^{-(\lambda_n + r)t})}{(\lambda_n + r)^2} \right].
\eea
For thermodynamically large system, \ie, $L \rightarrow \infty$, the sum in the above equation can be converted to an integral over $q=\frac{2{\pi}n}{L},$ and yields,
\bea
 \la J_d(t) \ra&=& (r_1 - r_2)\int_0^{2 \pi} \frac{dq}{2 \pi} \; \left[ \frac{t}{\lambda_q + r} -\frac{(1-e^{-(\lambda_q + r)t})}{(\lambda_q + r)^2} \right]\cr
 &=& (r_1 - r_2)\left[\frac{t}{\sqrt{r(4 + r)}}  - \frac{ (2+r)}{r(4+r)} + \int_0^{2 \pi}\frac{dq}{2 \pi} \; \frac{e^{-(\lambda_q + r)t}}{(\lambda_q + r)^2}\right]. \label{eq:Jdav}
\eea
In the long time regime, the exponential term decays and we have a linear temporal growth of the diffusive current,
\bea
 \la J_d(t) \ra=  \frac{(r_1 - r_2)t}{\sqrt{r(4 + r)}}=  {(2\alpha -1)}\sqrt{\frac r{4+r}}t. \label{eq:Jdav_lt}
\eea
It is to be noted that the  above expression has a very similar form to the long-time average current in case of single resetting (see Eq. (23) in Ref~\cite{sep-reset}); the dependence on the total resetting rate $r$ is the same in both the cases, but a prefactor $(r_1-r_2)$ arises in the presence of dichotomous resetting which allows the average current to become negative if $r_2 > r_1.$ For the special case when $r_1=r_2$, $\la \Jd(t)\ra$ vanishes. This feature is unique to dichotomous resetting and we will discuss it in more details in a later Section.

\begin{figure}[t]
 \centering
 \includegraphics[width=13.5 cm]{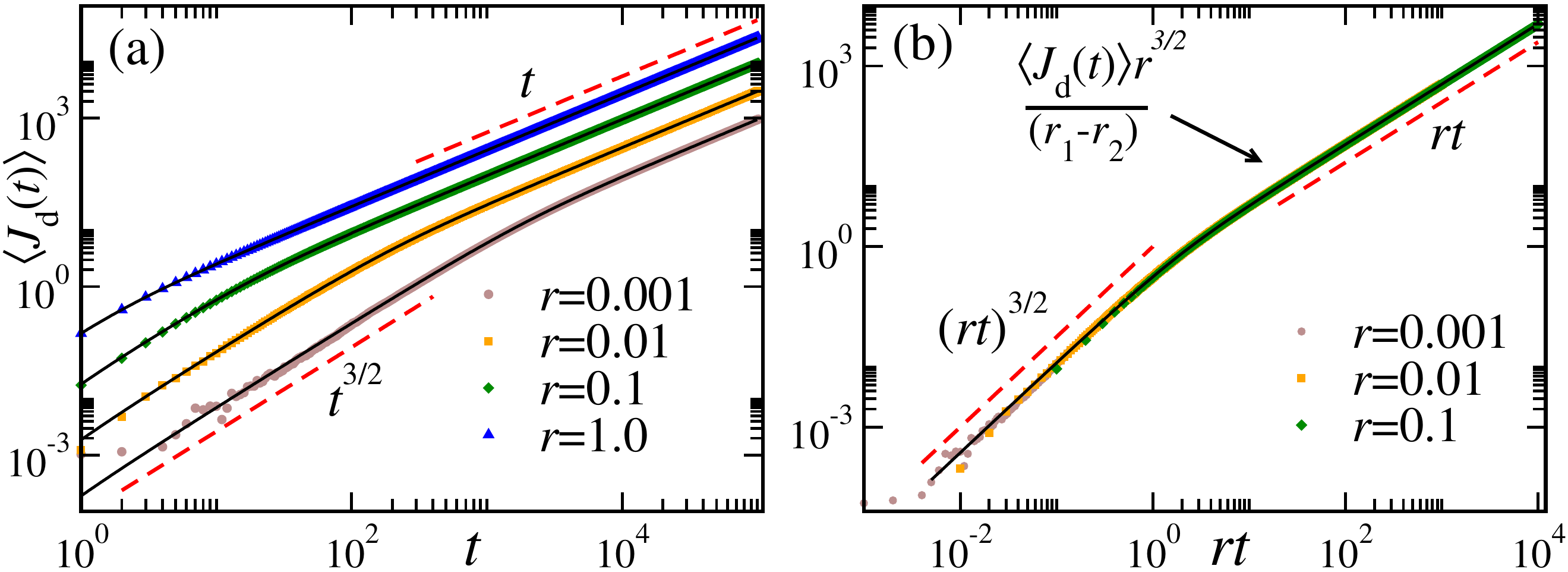}
\fl
 \caption{Diffusive current: (a) Plot of $\la \Jd (t)\ra$ as a function of time $t$ for different values of $r$ and $\alpha=0.8.$ The solid lines correspond to the exact results (see Eqs. \eref{eq:Jdav} and \eref{eq:Jdav_lt}) and the symbols correspond to the data obtained from numerical simulations. (b) Scaling collapse of the data in (a) according to Eq.~\eref{eq:Jd_smallr}. The solid line corresponds to the scaling function in Eq.~\eref{eq:Jd_smallr}. The numerical simulation is done on a lattice of size $L=1000$ and averaged over more than $10^7$ trajectories.}
 \label{fig:diff-cur}
\end{figure}

It is also interesting to investigate the short-time behaviour of the average current. Equation \eref{eq:Jdav} provides an exact expression which is valid at all times and can be used to compute $\la \Jd(t)\ra$ by performing the $q$-integral numerically. However, an alternative form for $\la \Jd(t)\ra,$ which lends itself more easily to numerical evaluation, can be derived using the renewal Equation \eref{eq:rho_renewal} for $\rho(x,t).$ 
Since the average instantaneous current $\la j_\id(s)\ra$ is the gradient of density across the central bond, we have a similar renewal equation for it,
\bea
\la j_\id(s)\ra = (r_1-r_2)\int_0^s \id s'~ \la j^1_0(s') \ra, \label{eq:j_renewal}
\eea
where $\la j_0^1(s) \ra$ denotes the average instantaneous current in the absence of resetting, starting from the configuration $\cal C_1.$ \footnote{The superscript $1$ indicates the initial configuration $\cal C_1$ and the subscript $0$ indicates the absence of resetting. We will use this convention again later.} Note that, there is no contribution from the initial condition as we have chosen the initial profile to be completely flat. 
The average instantaneous current $\la j_0^1(s) \ra$ is known exactly in terms of the Modified Bessel function of the first kind: $\la j_0^1(s) \ra = e^{-2 s} I_0(2s)$~\cite{sep-reset}. The average total current is then obtained by integrating Eq.~\eref{eq:j_renewal},
\bea
\la \Jd(t) \ra &=& (r_1-r_2)\int_0^t \id s \enskip \int_0^s \id s' \enskip e^{-(r+2)s'} I_0(2s') \cr
&=& (r_1-r_2)\int_0^t \id s ~ (t-s) \enskip e^{-(r+2)s} I_0(2s). \label{eq:Jd_2}
\eea
$\la \Jd(t) \ra$ at any time $t$ can be obtained by numerically evaluating the integral in the above equation; it can also be shown easily that Eq.~\eref{eq:Jd_2} is equivalent to Eq.~\eref{eq:Jdav}. 
Also, note that $\la J_d(r_1,r_2) \ra = -\la J_d(r_2,r_1) \ra$ at all times $t,$ hence it suffices to look at  $\la \Jd(t) \ra$ for $r_1 > r_2$ only. Figure \ref{fig:diff-cur}(a) shows plots of $\la \Jd(t) \ra$ obtained from Eq.~\eref{eq:Jd_2} for a set of values of $r$ (solid lines) for a fixed $\alpha,$ along with the same obtained from numerical simulations (symbols); a perfect agreement between the two sets verifies our analytical prediction.

Equation \eref{eq:Jd_2} can be used to derive an explicit expression for $\la \Jd(t) \ra$ for small values of $r \ll 1.$ In this case, the integrand is dominated by the large values of $s \gg r^{-1}.$ Using the asymptotic behaviour of $I_0(2x)$ for large $x$ (see Ref.~\cite{dlmf}, Eq. 10.40.1) and a variable transformation $u=rs,$ we get,
\bea
\la \Jd (t) \ra \simeq \frac{(r_1-r_2)}{2 \sqrt{\pi r^3}} \int_0^{rt}\id u ~ \frac{(rt-u)}{\sqrt{u}} e^{-u} .
\eea
The above integral can be computed exactly, and yields,
\bea
\la \Jd (t) \ra \simeq \frac{(r_1-r_2)}{\sqrt{r^3}} \left[\frac{e^{-rt}\sqrt{{rt}}}{2\sqrt{\pi}}+\frac{1}{4} (2rt-1) \enskip \textrm{erf}(\sqrt{rt})\right]. \label{eq:Jd_smallr}
\eea
Clearly, for large time $t \gg r^{-1},$ the above equation predicts $\la \Jd \ra \simeq (r_1 - r_2)t/2\sqrt{r},$ which is consistent with Eq. \eref{eq:Jdav_lt} in the small $r$ limit. One can also extract the short-time behaviour of the average current from Eq.~\eref{eq:Jd_smallr} and it turns out that for $t \ll r^{-1},$ the average current shows a super-linear growth,
\bea
\la \Jd(t) \ra \simeq \frac{3(r_1-r_2)}{2 \sqrt{\pi}} t^{3/2} +\cal O(t^{5/2}).
\eea
It should be emphasized that the short-time behaviour depends on the specific initial condition considered. Depending on the choice of initial configuration, the current can show very different short-time behaviour --- it is easy to see that, while starting with $\cal C_1$ and $\cal C_2$ with equal probabilities leads to the $t^{3/2}$ growth, starting with $\cal C_1$ only leads to a $\sim \sqrt{t}$ behaviour.

We will conclude the discussion about the average diffusive current with one final comment. From Eq.~\eref{eq:Jd_smallr}, it appears that $\la \Jd (t)\ra r^{3/2}/(r_1-r_2)$ depends only on $rt,$ and not on $r_1, r_2$ separately. 
Figure \ref{fig:diff-cur}(b) shows a plot of $\la \Jd (t)\ra r^{3/2}/(r_1-r_2)$ as a function of $rt$ for different small values of $r;$ the collapsed curve is compared with the scaling function predicted by Eq.~\eref{eq:Jd_smallr} (solid line). \\

\noindent {\bf Fluctuations of $\Jd$:} To characterize the behaviour of the diffusive current, it is also important to understand the nature of its fluctuations. 
To this end, we investigate the higher moments of $\Jd,$ starting with its variance. We use the method introduced in Ref.~\cite{sep-reset} and note that the net diffusive current along any trajectory in the duration $[0,t]$ can be written as a sum of the currents in the intervals between consecutive resetting events. Let us assume that there are $n$ resetting events (irrespective of the configuration to which the system is reset) during the interval $[0,t]$ and let $t_i$ denote the interval between the $(i-1)^{th}$ and $i^{th}$ events. The probability of such a trajectory is given by,
\bea
\cal P_n(\{ t_i \}) = r^n e^{-r \sum_{i=1}^{n+1} t_i}, \label{eq:Pn_ti}
\eea
where $t= \sum_{i=1}^{n+1} t_i.$ Here, $t_1$ is the time before the first resetting event and  $t_{n+1}$ denotes the time between the last resetting and final time $t.$  Along this trajectory, the total diffusive current is,
\bea
\Jd = \sum_{i=1}^{n+1} J_0(t_i), \label{eq:Jd_sum}
\eea
where $J_0(t_i)$ denotes the net hopping current during the interval $t_i,$ in the absence of resetting, but starting from $\cal C_1,$ or $\cal C_2,$ depending on the $(i-1)^{th}$ resetting event. Let us recall that, the resetting to $\cal C_1,$ respectively  $\cal C_2,$ occurs with probability $\alpha = \frac{r_1}{r_1+r_2},$ respectively $1-\alpha= \frac{r_2}{r_1+r_2}$
Then, the probability that the value of the diffusive current is $J_i \equiv J_0(t_i)$ during an interval $t_i$ is given by,
\bea
P(J_i,t_i) = \alpha P_0^1(J_i,t_i) +(1-\alpha)P_0^2(J_i,t_i) \quad \textrm{for} \; 2 \le i \le n+1,
\eea
where $P_0^1(J_0,\tau)$ (respectively, $P_0^2(J_0,\tau)$) denotes the probability that the diffusive current will have a value $J_0$ during the interval $\tau$ starting from $\cal C_1$ (respectively, $\cal C_2$) in the absence of resetting. However, before the first resetting event, \ie, for the time interval $[0,t_1]$, we have,
\bea
P(J,t_1) = \frac {1}{2} \left[P_0^1(J,t_1) + P_0^2(J,t_1)\right],\n
\eea
as we start from $\cal C_1$ and $\cal C_2$ with equal probability.

We can now write the probability that, in the presence of dichotomous resetting, the total diffusive current has a values $\Jd$ at time $t,$ 
\bea
\fl P(J_d,t)= \sum_{n=0}^\infty \int_0^t \prod_{i=1}^{n+1}\id t_i ~ \cal P_n(\{ t_i \}) \, \delta (t- \sum_{i=1}^{n+1} t_i)  \,  \int  \prod_{i=1}^{n+1}\id J_i \, P(J_i,t_i) \, \delta (J_d- \sum_i J_i), \label{eq:P_jd} 
\eea
where we have used the fact that the hopping currents $J_i$ in the intervals $t_i$ are independent of each other.
To circumvent the constraints presented by the $\delta$-functions, it is convenient to work with the Laplace transform of the moment generating function $ \la e^{\lambda J_d} \ra $ with respect to time,
\bea
Q(s,\lambda) = \cal {L}_{t \rightarrow s}[ \la e^{\lambda J_d} \ra ] = \int_0^{\infty} \id t \, e^{-st}  \int \id \Jd \, e^{\lambda J_d} P(J_d,t). \label{eq:Q_s}
\eea
\begin{figure}[t]
 \centering
 \includegraphics[width=9 cm]{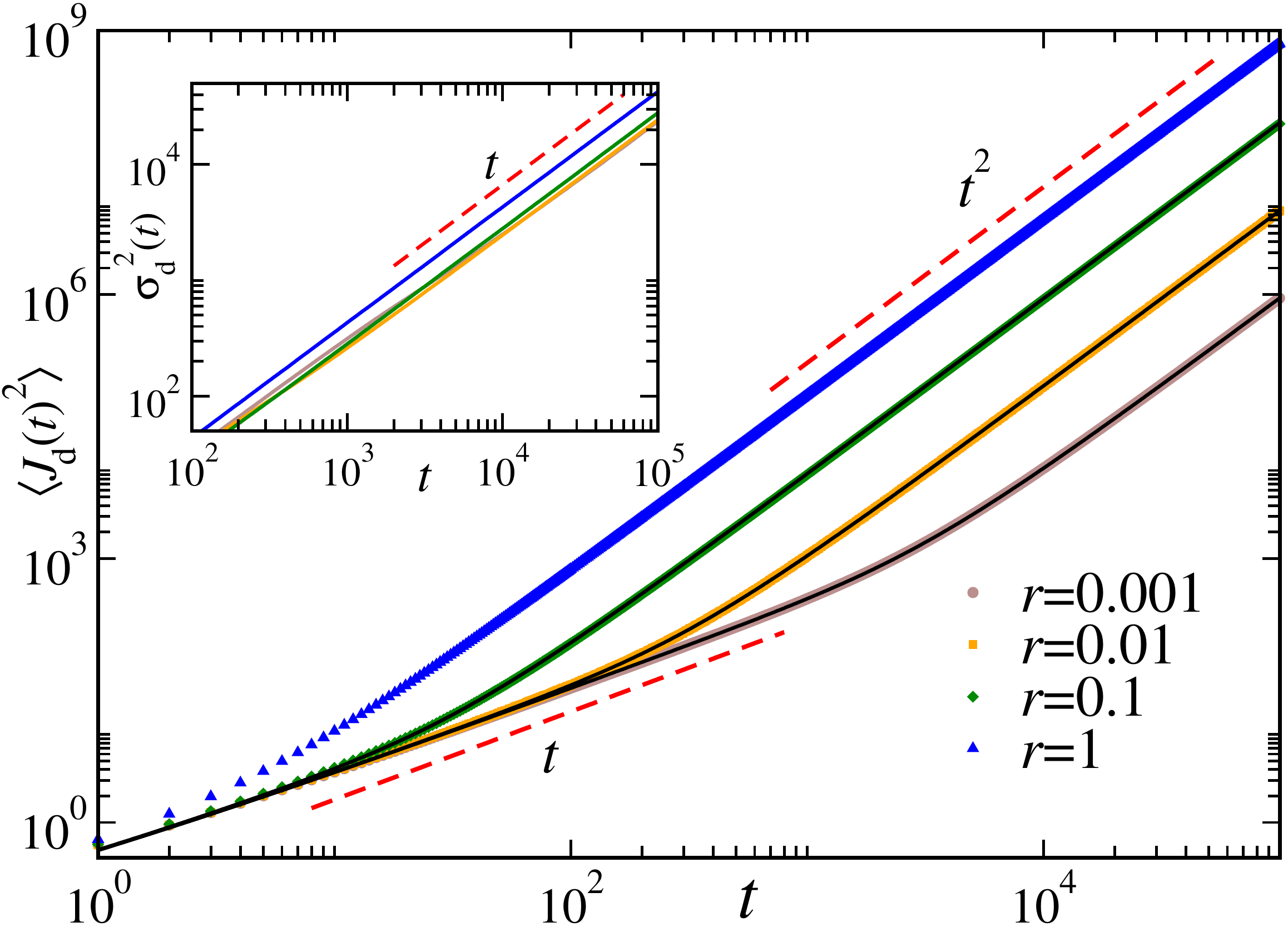}
 \caption{The second moment $\la \Jd(t)^2 \ra$ of the diffusive current as a function of time $t$ for different values of $r$ and $\alpha_1=0.8.$ The symbols correspond to the data obtained from numerical simulations and the black solid lines refer the analytical result for small $r$ [see Eq.~\eref{eq:Jd2}]. The inset shows the corresponding plot for the variance $\sigma_\id^2(t).$ The system size $L=1000$ for the simulations. }
 \label{fig:varJ}
\end{figure}
Using Eq.~\eref{eq:P_jd} and performing the integrals over $\Jd$ and $t,$ we get,
\bea
Q(s,\lambda) &=& \sum_{n=0}^{\infty} r^n  \int_0^{\infty}  \prod_{i=1}^{n+1}dt_i \enskip e^{-(r+s)  \sum_i t_i} \enskip  \int \displaystyle \prod_{i=1}^{n+1}dJ_i \enskip e^{\lambda  \sum_i J_i} \enskip P(J_i,t_i) \cr
&=& h_{\frac 12}(s,\lambda)\sum_{n=0}^{\infty} r^n  \enskip h_{\alpha}(s,\lambda)^n, \label{eq:Q_s_sum}
\eea
where, 
\bea
\fl h_{\alpha}(s,\lambda) &=& \int_0^{\infty} d\tau \enskip e^{-(r+s)\tau} \enskip \left[\alpha \int \id J_0\enskip e^{\lambda J_0} P_0^1(J_0,\tau)+ (1-\alpha) \int \id J_0 \enskip e^{\lambda J_0} P_0^2(J_0,\tau)\right . \label{eq:h-def}
\eea
Now, let us recall that, the configurations $\cal C_1$ and $\cal C_2$ are complementary to each other --- during some time interval $\tau,$ corresponding to each trajectory starting from $\cal C_1,$ with net current $J_0,$ there exists a trajectory starting from $\cal C_2,$ which yields net current $-J_0.$ In other words, $P_0^2(J_0,\tau) = P_0^1(-J_0,\tau).$ Using this relation, Eq.~\eref{eq:h-def} yields,
\bea
h_\alpha(s,\lambda) = \alpha h(s,\lambda) + (1-\alpha) h(s,-\lambda),\label{eq:h_slam}
\eea
where, for the sake of notational convenience, we have denoted,
\bea
h(s,\lambda)= \int_0^{\infty} d\tau \enskip e^{-(r+s)\tau} \int \id J_0\enskip e^{\lambda J_0} P_0^1(J_0,\tau).  \label{eq:h_s}
\eea 
Using Eqs.~\eref{eq:h_slam} and \eref{eq:Q_s_sum}, and performing the sum over $n,$
we finally have,
\bea
Q(s,\lambda) = \frac{h(s,\lambda)+h(s,-\lambda)}{2[1- r\{\alpha h(s,\lambda)+(1-\alpha)h(s,-\lambda)\}]}.\label{eq:Q_sl}
\eea
To calculate $h(s,\lambda)$ exactly one needs $P_0^1(J_0,\tau),$ the current distribution in the absence of resetting, which, unfortunately, is not known for arbitrary values of $\tau.$
However, following the approach used in Ref.~\cite{sep-reset}, we can compute $h(s,\lambda)$ for small values of $r$ and $s$ using the large-time moment generating function of $J_0(\tau),$ derived in Ref.~\cite{Derrida1}. 
In particular, for the initial configuration $\cal C_1$ it has been shown in Ref.~\cite{Derrida1,sep-reset} that, for large values of $\tau,$ 
\bea
\int \id J_0~ e^{\lambda J_0} P^1_0(J_0,\tau) \simeq e^{\sqrt{\tau}F(\lambda)}, \quad \textrm{with} \enskip F(\lambda)=-\frac{1}{\sqrt{\pi}} \textrm{Li}_{3/2}(1-e^{\lambda}), \label{eq:eJ0}
\eea
where $\textrm{Li}_{\alpha}(z)$ denotes the Poly-Logarithm function (see Ref.~\cite{dlmf}, Eq.~25.12.10).\\
Substituting Eq. \eref{eq:eJ0} in Eq. \eref{eq:h_s} and evaluating the integral, we get, for small $r,s,$
\bea
h(s,\lambda) \simeq \frac{1}{r+s}\left[1+\frac{\sqrt{\pi}F(\lambda)}{2 \sqrt{r+s}}e^{\frac{F(\lambda)^2}{4(r+s)}}\left(1+ \textrm{erf}\left[\frac{F(\lambda)}{2 \sqrt{r+s}}\right]\right)\right].\label{eq:h_sl}
\eea 
Now, we can extract the Laplace transform of any moment of $\Jd$  using Eq.~\eref{eq:h_sl} along with Eq.~\eref{eq:Q_sl}. First, we have,
\bea
\cal L_{t \rightarrow s} [\la J_d(t)\ra] = \frac{\id}{\id\lambda} Q(s,\lambda) \Bigg|_{\lambda=0} = \frac{r_1-r_2}{2 s^2 \sqrt{r+s}}.
\eea
The average current is obtained by taking the inverse Laplace transform,
\bea
\la J_d(t)\ra &=&\cal L_{s \rightarrow t}^{-1}\left[ \frac{r_1-r_2}{2 s^2 \sqrt{r+s}} \right] \cr
&=&\frac{(r_1-r_2)}{\sqrt{r^3}} \left[\frac{e^{-rt}\sqrt{{rt}}}{2\sqrt{\pi}}+\frac{1}{4} (2rt-1) \enskip \textrm {erf}(\sqrt{rt})\right].\label{eq:Jav_Lt}
\eea
Note that, as expected, the above equation is identical to Eq.~\eref{eq:Jd_smallr}, which is also valid for small values of $r$ and large $t,$ albeit obtained using a different method.

Next, we calculate the second moment $\la \Jd(t)^2\ra.$ The corresponding Laplace transform is obtained from Eq.~\eref{eq:Q_sl},
\bea
\cal L_{t \rightarrow s} [\la J_d^2(t)\ra] &=& \frac{\id^2}{\id \lambda^2} Q(s,\lambda) \Bigg|_{\lambda=0} \cr
&=& \frac{1}{\pi s^2}+\frac{(r_1-r_2)^2}{2s^3\sqrt{r+s}}+\frac{b\sqrt{r+s}}{2s^2},
\eea
where $b= 1- \frac 1{\sqrt{2}}.$ Fortunately, we can invert the Laplace transform exactly to obtain the second moment of the diffusive current for small $r$ and large values of $t,$
\bea
 \la \Jd^2(t) \ra &=& \frac{(r_1-r_2)^2}{2 r^3} + e^{-rt} \bigg[ \frac b 2 \sqrt{\frac t \pi}- \frac{(r_1-r_2)^2}{2 r^3}\bigg]  + t \bigg[\frac 1 \pi - \frac{(r_1-r_2)^2}{2r^2}\bigg]\cr
&& + \frac{(r_1-r_2)^2 t^2}{4r} + \frac{b}{4 \sqrt{r}}(1+2rt) \enskip \text{erf}(\sqrt{rt}). \label{eq:Jd2}
\eea
In the long-time limit $\la \Jd^2(t)\ra$ shows a quadratic behaviour for any $r_1 \ne r_2.$ The variance $\sigma_d^2 =\la \Jd^2(t) \ra - \la \Jd(t) \ra^2 $ can be obtained using Eqs.~\eref{eq:Jav_Lt} and \eref{eq:Jd2}. In particular, in the long-time limit, the variance increases linearly with time,
\bea
\sigma_\id^2(t) \simeq t \bigg [\frac 1 \pi +\frac {\sqrt{r}}2 \bigg(1 - \frac 1 {\sqrt{2}}\bigg)- \frac{(r_1-r_2)^2}{4 r^2}\bigg].\label{eq:sigma}
\eea
Figure \eref{fig:varJ} shows a plot of $\la \Jd^2(t)\ra$ {\it vs} time for a set of values of $r.$ The inset shows the corresponding variance $\sigma_\id^2(t)$ in the long-time regime.  

\begin{figure}[t]
 \centering
 \includegraphics[width=10 cm]{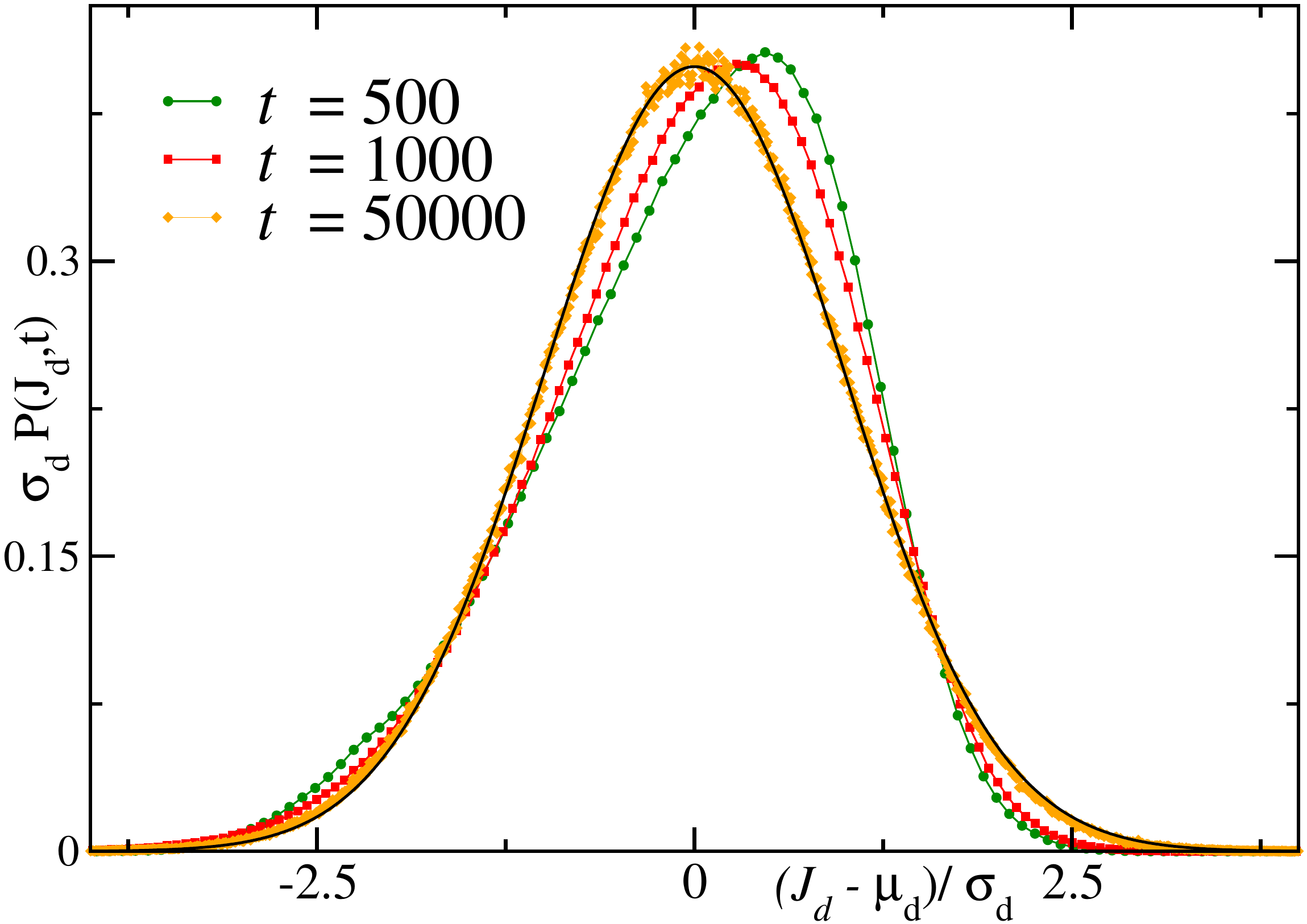}
 \caption{Probability distribution of the diffusive current. The symbols correspond to the data obtained from numerical simulations and the black solid line refers to the analytical result for large $t$ [see Eq.~\eref{eq:PJ_d_lt}]. Here, $r_1=0.008$ and $\alpha_1=0.8$. The system size is $L=1000$. }
 \label{fig:P_Jd}
\end{figure}

To understand the nature of the fluctuation in more detail, next we explore the probability distribution $P(\Jd,t)$. As noted in Eq. \eref{eq:Jd_sum}, the net diffusive current $\Jd$ is a sum of current $J_0(t_i)$ during intervals $t_i$ between two consecutive resetting events. These $J_0(t_i)$ are completely independent of each other, and central limit theorem predicts that, when the number of resetting events is large, \ie, for $rt \gg 1,$  the typical distribution of the sum should be a Gaussian, 
\bea
P(\Jd,t) = \frac 1{\sqrt{2 \pi \sigma_\id^2(t)}} \exp \bigg[- \frac{(\Jd - \la \Jd (t)\ra)^2}{2 \sigma_\id^2(t)}\bigg]. \label{eq:PJ_d_lt}
\eea
It should be noted that, even for the case of resetting to a single configuration, the diffusive current shows a Gaussian behaviour in the long-time regime, although the mean and the variance are very different in that case. 
Figure \eref{fig:P_Jd} shows plots of $P(\Jd,t)$ for different values of $t$ which shows a very good agreement with the Gaussian (solid line) in the large time limit. 

The argument used above to predict the Gaussian nature of $P(\Jd,t)$ relies crucially on the central limit theorem, which holds true only when there are large number of resetting events, \ie, $t \gg r^{-1}.$  It is interesting to investigate how the distribution approaches the Gaussian limit.  To understand this approach, we compute the skewness and the kurtosis of the distribution as a function of time.  The skewness measures the `asymmetry' in the distribution, and is defined as,
\bea
\gamma= \frac{\la (\Jd - \la \Jd \ra)^3\ra}{\sigma_\id^3} = \frac{\la J_d^3 \ra - 3 \mu_\id \sigma_\id^2- \mu_\id^3}{\sigma_\id^3}, \label{eq:skewness_gamma}
\eea
where we have used $\mu_\id \equiv \la \Jd \ra$ for notational brevity. A Gaussian distribution is symmetric around the mean and the skewness vanishes. A positive (negative) value of the skewness indicates a tail towards the right (left) side of the distribution. On the other hand, the kurtosis measures the `peakedness' of a distribution and is defined as,
\bea
\kappa = \frac{\la (\Jd - \la \Jd \ra)^4\ra}{\sigma_\id^4} = \frac{\la J_d^4 \ra - 4 \mu_\id \la J_d^3 \ra +6 \mu_\id^2 \sigma_\id^2 +3 \mu_\id^4}{\sigma_\id^4}, \label{eq:kurt}
\eea
For a Gaussian distribution $\kappa =3;$ if $\kappa >3$ it indicates a heavy-tailed distribution compared to a Gaussian one whereas $\kappa <3$ indicates a more `peaked' distribution.

To calculate $\gamma$ and $\kappa$ for $P(\Jd, t)$ we need the third and fourth moments which can be calculated in a straightforward manner from Eq.~\eref{eq:Q_sl}. The Laplace transform of the third moment is given by,
\bea
\fl \quad \cal L_{t \rightarrow s} [\la \Jd^3(t)\ra] = \frac{\id^3}{\id \lambda^3} Q(s,\lambda) \Bigg|_{\lambda=0}\cr
= \frac{(r_1-r_2)}{12 \pi s^4 (r+s)^{3/2}}\bigg[9 \pi \bigg( (r_1-r_2)^2+brs(1+2s)(\sqrt{r+s})\bigg)\cr
\qquad \qquad \qquad \quad~ 9s(4r+3s+4bs)-c_1 \pi s^2(r+s)\bigg],
\eea
where $b=1- \frac 1{\sqrt{2}}$ and $c_1=-6+9\sqrt{2}-4 \sqrt{3}.$ Similarly,
\bea
\fl \cal L_{t \rightarrow s} [\la \Jd^4(t)\ra] = \frac{\id^4}{\id \lambda^4} Q(s,\lambda) \Bigg|_{\lambda=0}\cr
= \frac{3(r_1-r_2)^4}{2 s^5(r+s)^2} 
+\frac{3(r_1-r_2)^2}{2 \pi s^4(r+s)^2}\bigg[2(3r+2s) +b \sqrt{r+s}((8+\pi)s+3 \pi r)\bigg]\cr
+\frac 1{3 \pi^2 s^3(r+s)}[6(3r+s)-c_1\pi^2 (r_1-r_2)^2]+\frac{3 b (4r+3s)}{2 \pi s^3\sqrt{r+s}}\cr
-\frac 1{12 s^3}[3\pi c_2 s \sqrt{r+s} + 8 c_1 s - 18b^2(\pi r +2s)],
\eea
where, as before, $b=1- \frac 1{\sqrt{2}}$ and $c_1=-6+9\sqrt{2}-4 \sqrt{3},$  $c_2=4+7\sqrt{2}-8 \sqrt{3}.$
The Laplace transforms can be inverted exactly in both cases to find $\la \Jd^3(t)\ra$ and $\la \Jd^4(t)\ra$. However, the expressions are rather long and complicated, and as we are interested in the approach to the Gaussian, it suffices to provide the expression in the long-time limit only. The third moment is proportional to $(r_1-r_2),$ and grows as $t^3$ at large times,
\bea
\fl \langle J_d^3(t)\rangle = (r_1-r_2) \Bigg[\frac{(r_1-r_2)^2t^3}{8 r^{3 \over 2}}+ \frac{3 t^2}{16 \pi r^{5\over 2}}\left(4b \pi r^{5\over 2}+8 r^2-3\pi(r_1-r_2)^2\right)\cr
+\frac{t}{96 \pi r^{7\over 2}}\bigg(135\pi (r_1-r_2)^2- 72 r^2(3+b(\pi -4)\sqrt{r})+\pi c_1 r \bigg)\Bigg]+ \cal{O}(1). \label{eq:Jd3_lt}
\eea
The fourth moment grows as $t^4$ for large $t,$
\bea
\fl \la \Jd(t)^4 \ra = \frac{(r_1-r_2)^4 t^4}{16 r^2} + \frac{(r_1-r_2)^2 t^3}{4 \pi r^3} \bigg[6r^2+3 b \pi r^{5/2}- 2 \pi (r_1-r_2)^2 \bigg] \cr
+ t^2 \bigg[3 \left(\frac 1 \pi + \frac {b \sqrt{r}}2  \right)^2 +  \frac{9(r_1-r_2)^4}{4r^4}\cr
 -(r_1-r_2)^2 \left( \frac 6{\pi r^2}-\frac{3 b(16-7\pi)}{8 \pi r^{3 \over 2}}+\frac {c_1}{6r} \right) \bigg] + {\cal O}(t). \label{eq:Jd4_lt}
\eea
\begin{figure}[t]
 \centering
 \includegraphics[width=12.8 cm]{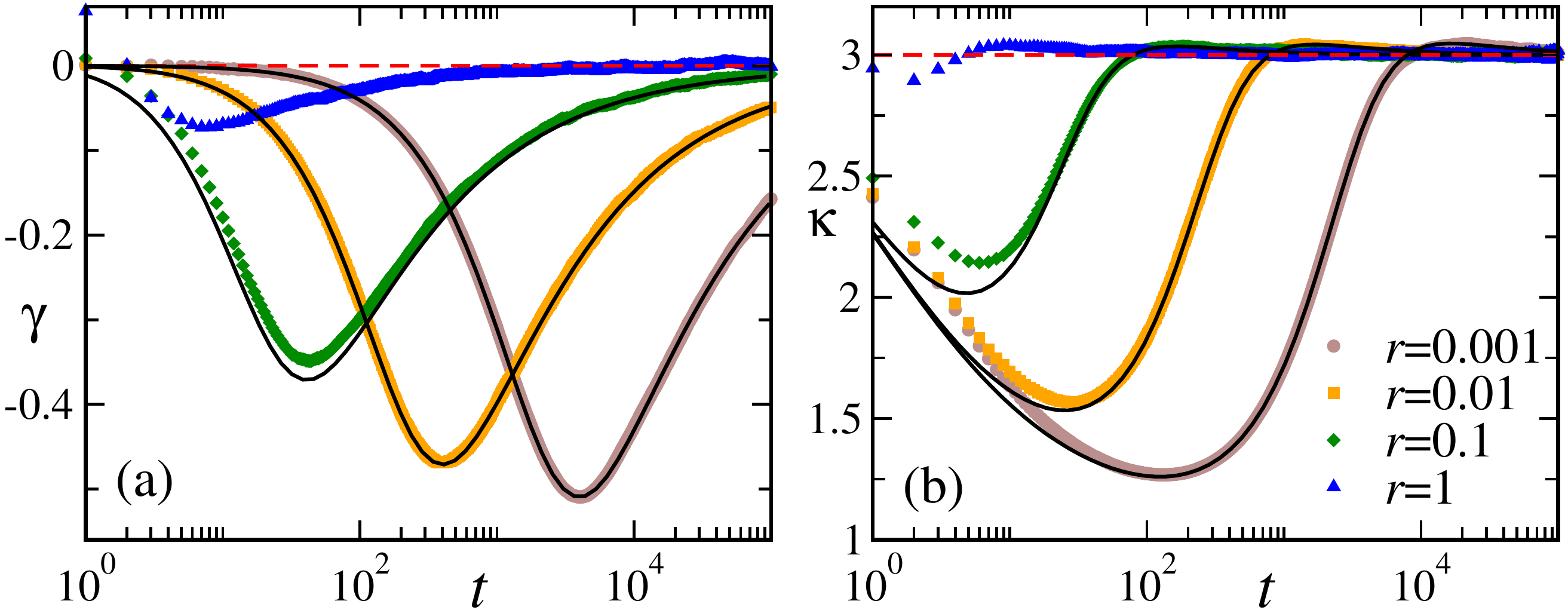}
 \caption{Skewness $\gamma$ (a) and kurtosis $\kappa$ (b) as a function of time $t$ for different values of $r$ and $\alpha=0.8.$ The symbols correspond to the data from the numerical simulations and the black solid lines correspond to the analytical predictions (see Eqs. \eref{eq:Jd3_lt} and \eref{eq:Jd4_lt}). The system size used for the numerical simulations is $L=1000.$}
 \label{fig:skew_kurt}
\end{figure}

From the above expressions it is straightforward to see that the third central moment $\la (\Jd - \la \Jd \ra)^3\ra \sim t$ at long-times. Hence, clearly, the skewness $\gamma$ vanishes as $\sim 1/\sqrt{t}$, indicating that the current distribution becomes symmetric at late times. The exact time-dependent $\gamma$ can be evaluated using Eqs. \eref{eq:Jav_Lt}, \eref{eq:Jd2}, \eref{eq:skewness_gamma} \eref{eq:Jd3_lt}. This is shown in Fig.~\ref{fig:skew_kurt}(a) where the analytical prediction is compared to the data from numerical simulation. Clearly, $\gamma$ vanishes for $t \gg r^{-1},$ whereas in the regime $ 1 \ll t \ll r^{-1}$ it has a negative value indicating an extended tail towards smaller values of $\Jd.$

The kurtosis $\kappa$ can also be calculated using the above equations.  Figure \ref{fig:skew_kurt}(b) compares the analytical prediction  of $\kappa$ with the same obtained from numerical simulations for different values of $r.$ As expected, the two agree very well for $t \gg 1.$ The figure shows that, starting from zero, the kurtosis first decreases, reaches a minimum and then increases and becomes larger than $3$ for a brief regime before approaching $\kappa=3$ for large times $t \gg r^{-1}.$ While the limit $\kappa = 3$  signifies the Gaussian nature of $P(\Jd,t)$ at large times, $\kappa <3$ at short-times indicates a sharply peaked distribution compared to a Gaussian one. To understand the approach to the Gaussian limit, we calculate the limiting behaviour of $\kappa$ using the expressions for the moments,
\bea
\lim_{t \to \infty} \kappa_3 = 3 + \frac A{t}, \label{eq:kurt_limit}
\eea
where $A$ is a non-zero constant which depends on $r_1,r_2$. Similar to the skewness, $\kappa$ also  approaches the Gaussian limit in an algebraic manner, albeit with a larger exponent $-1.$

%

\section{Zero-current state}\label{sec:zero_cur}

A special scenario emerges when the two resetting rates are equal, \ie, $r_1=r_2=\frac r 2.$ In this case, the resettings to $\cal C_1$ and $\cal C_2$ occur equally frequently, and consequently, the average density profile becomes flat in the stationary state and there is no diffusive particle current flowing through the system.  This is reminiscent of the equilibrium SEP on a periodic lattice, which also has these two features. However, in the presence of the dichotomous resetting, there is no detailed balance, and the system remains far away from equilibrium. It is then interesting to ask how one can characterize this zero-current nonequilibrium state and how is it different than the equilibrium state of ordinary SEP.


In this section we investigate this question and illustrate various aspects of the ZCS which distinguish it from the equilibrium SEP. Following the results presented in Sec.~\ref{sec:current}, we show that the temporal behaviour of the current fluctuations are different in the two cases. It also turns out that this ZCS shows non-trivial spatial and temporal correlations which are also very different than ordinary SEP. Moreover, we explore the response of this stationary ZCS to an external perturbation and show that the susceptibility in this case is drastically different  from the same in the equilibrium SEP.

%

\subsection{Current Fluctuations}\label{sec:zcs_current}

The first significant difference between the equilibrium SEP and the ZCS in the presence of dichotomous resetting shows up in the fluctuation of the diffusive current. 
The temporal behaviour of the equilibrium fluctuations of the current  has been studied in Ref.~\cite{Derrida1}. This corresponds to the case where the initial densities in the left and right half of the system are equal. Adapting their result to our case ($\rho=1/2$) we have, in the long time limit,
\bea
\la J_0^2(t)\ra \simeq \frac 12 \sqrt{\frac{t}\pi},\cr
\la J_0^4(t)\ra \simeq \frac {3t}{4 \pi} + \frac 18 \sqrt{\frac{t}\pi}(4-3\sqrt{2}),\label{eq:J02_J04}
\eea
while, of course, the odd moments vanish. Moreover, the distribution of the current $J_0(t)$ becomes  Gaussian at long-times;  the approach to the Gaussian is characterized by an algebraic decay of the kurtosis at long times,
\bea
\lim_{t \to \infty} \kappa = 3 - (3- 2 \sqrt{2})\sqrt{\frac{\pi}{2t}} + O\left(\frac 1t \right). 
\eea

For the ZCS, the moments of the diffusive current can be obtained by putting $r_1=r_2$ in Eqs.~\eref{eq:Jd2} and \eref{eq:Jd4_lt}  and we have, in the long time limit,
\bea
\la \Jd^2(t) \ra &\simeq & t \left(\frac 1 \pi + \frac{b \sqrt r}2 \right), \cr
\la \Jd^4(t) \ra &\simeq & 3 t^2 \left(\frac 1 \pi + \frac {b \sqrt r}2 \right)^2.
\eea
Clearly, the fluctuations grow much faster compared to equilibrium case. Note that, the moments grow slower compared to the case $r_1 \ne r_2$ where $\la \Jd (t)^2 \ra \sim t^2$ and $\la \Jd(t)^4 \ra \sim t^4.$
All the odd moments vanish, of course, and the skewness remains zero at all times. This is expected, as we are starting from a symmetric initial condition and $r_1=r_2$ does not introduce any directional bias. Kurtosis, on the other hand,  approaches the Gaussian value $3$ in the long time limit as $t^{-1}$; see Eq.~\eref{eq:kurt_limit}. Hence, the approach to a Gaussian distribution for the ZCS is much faster compared to the equilibrium case.



\subsection{Configuration weights}\label{sec:zcs_config}

As mentioned already, the ZCS is characterized by a flat stationary profile. However, the weights of different configurations which contribute to this flat profile need not be same. Let us recall that, for ordinary SEP on a periodic lattice, each configuration becomes equally likely in the equilibrium state. In presence of the dichotomous resetting with equal rates $r_1=r_2,$ the stationary probability of any configuration $\cal C$ is obtained by taking the $t \to \infty$ limit in Eq.~\eref{eq:PC_sol2},
\bea
\cal P_{\textrm st}(\cal C) = \frac r2 \int_0^{\infty} \id s ~ e^{-r s}~ \bigg [\cal P_0(\cal C,s | \cal C_1,0)+\cal P_0(\cal C,s | \cal C_2,0) \bigg],
\eea
where $\cal P_0$ denotes the configuration probabilities in the absence of resetting. 
For a large resetting rate $r,$ the integration is dominated by the contribution from $s \ll r^{-1},$ and hence $\cal P_{\textrm st}(\cal C)$ will be large for those configurations $\cal C$ 
for which either $\cal P_0(\cal C,s | \cal C_1,0)$ or $\cal P_0(\cal C,s | \cal C_2,0)$ is large for small $s,$ \ie, configurations which are `dynamically close' to $\cal C_1$ and  $\cal C_2.$  On the other hand, for smaller values of $r,$ the integration is dominated by large values of $s,$ and $\cal P_{\textrm st}(\cal C)$ would have significant contributions for configurations `far' from $\cal C_1$ and  $\cal C_2.$

To illustrate this point we take the simplest example of a lattice of size $L=4$ and $L/2=2$ particles. In this case, there are 6 possible configurations including the resetting configurations $\cal C_1=1100$ and $C_2=0011.$ The stationary weights of the configurations can be calculated exactly, and yields, 
$\cal P_{\textrm{st}}(\cal C_1)=P_{\textrm{st}}(\cal C_2)= \frac{r(r+6)+4}{2(r+2)(r+6)}.$ Moreover, $\cal P_{\textrm{st}}(1010)=\cal P_{\textrm{st}}(0101) = \frac 1{r+6}$ and $\cal P_{\textrm{st}}(1001) = \cal P_{\textrm{st}}(0110)= \frac 2{(r+2)(r+6)}.$
Of course, when $r=0,$ all the six configurations occur with equal weight. For any non-zero $r$ the resetting configurations $\cal C_1$ and $\cal C_2$ have the highest stationary probability. For large values of $r,$ the weight of the configurations $1010$ and $0101,$ which can be reached from $\cal C_1$ or $\cal C_2$ by one hop (\ie, dynamically close) vary as $1/r$ where as the weight of the configurations $1001$ and $0110$ (which are further away from the resetting configuration) decay as $1/r^2.$ On the other hand, for small $r$ all the configurations have  comparable, although different, stationary weights.

\subsection{Spatial correlations} \label{subsec:spat_correl}

It is interesting to investigate the spatial correlation of the SEP in the presence of the dichotomous resetting. Ordinary SEP, in the limit of thermodynamically large system size, has a product measure  stationary state, so that the connected correlations vanish. In the presence of resetting, however, one can expect non-trivial spatial correlations, even for $r_1=r_2.$  In this section we explore the behaviour of the two point correlation $C_{x,y} =\la s_x s_y \ra$ in the presence of the dichotomous resetting.

For ordinary SEP, in equilibrium,  all the configurations are equally likely. In particular, for a half-filled system of size $L,$ each configuration has a probability $^LC_{L/2}$ to occur in the stationary (equilibrium) state.  Correspondingly, the stationary two-point correlation for any finite system size $L$ is given by, 
\bea
C_{x,y} = \frac{(L-2)}{4(L-1)}, \label{eq:cxy_eq}
\eea
which is independent of $x$ and $y$ and converges to $\rho^2=1/4$ for a thermodynamically large system. 

\begin{figure}
\includegraphics[width=8 cm]{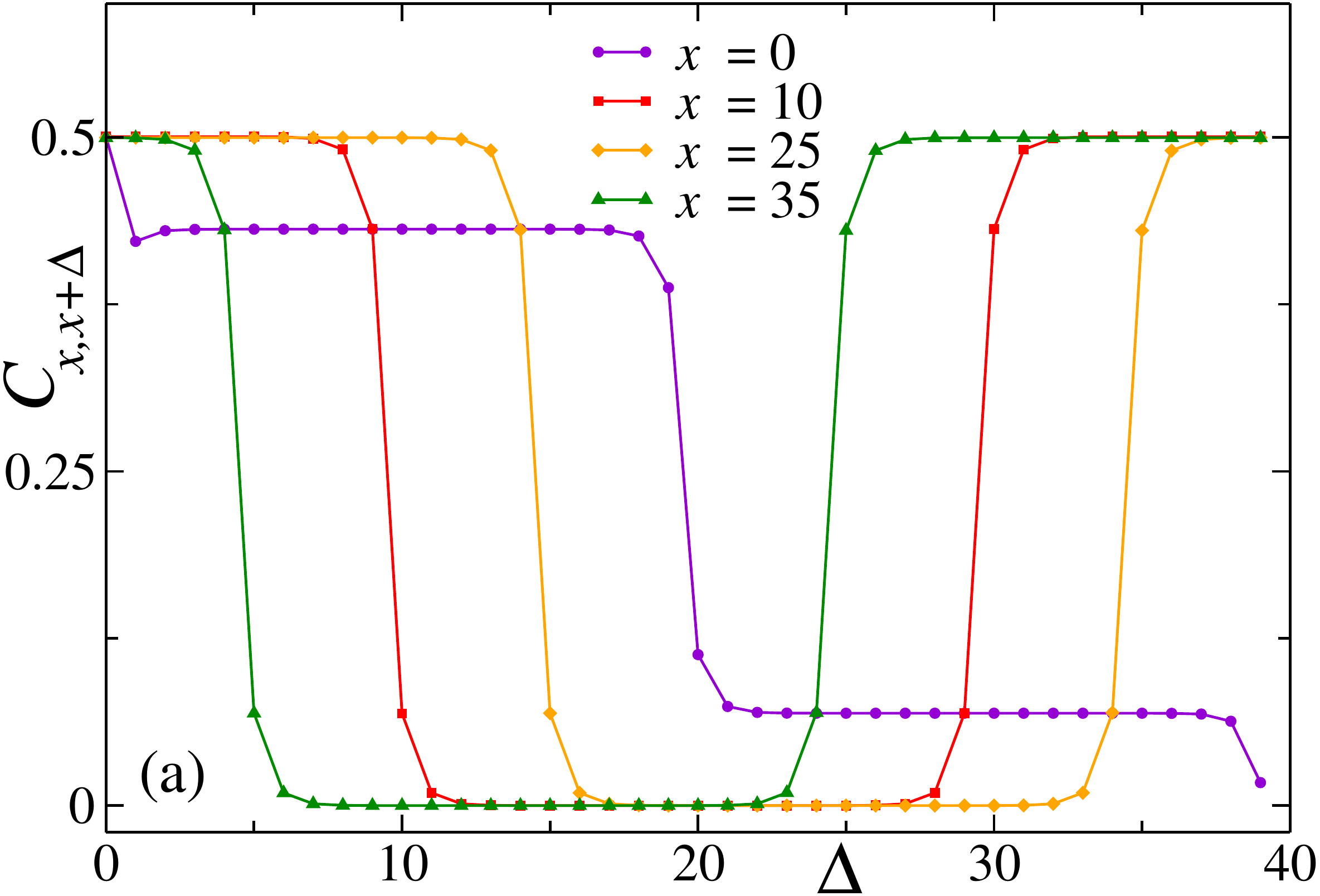}  \includegraphics[width=8 cm]{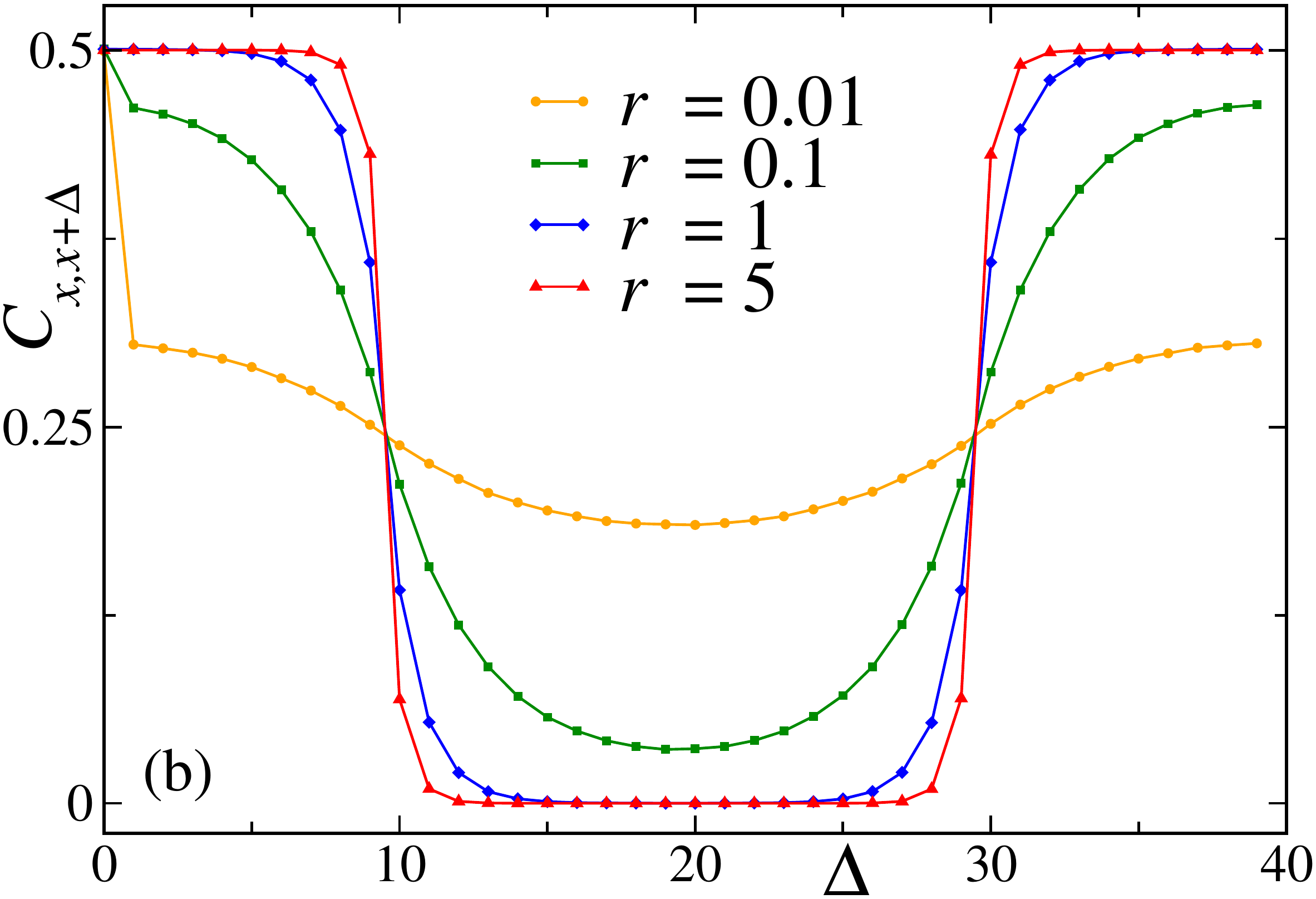}  
 \caption{Plot of the two-point correlation $C_{x, x+ \Delta}$ as a function of $\Delta$ for (a) a fixed (large) value of $r=5$ and different values of $x$ and (b) for fixed $x=10$ and different values of $r.$  The system size $L=40$ here.}\label{fig:corr_cxy}
\end{figure}

In the presence of the resetting, it is straightforward to write a renewal equation for $C_{x,y}(t)= \sum_{\cal C} s_x(t) s_y(t) \cal P(\cal C,t| \cal C_0,0)$ using Eq.~\eref{eq:PC_sol2}. We are particularly interested in the 
case $r_1=r_2$ and for the stationary correlation  which is obtained by taking the limit $t \to \infty,$ 
\bea
 C_{x,y} = \frac r2 \int_0^\infty \id s~e^{-r s} \left [\la s_x(s) s_y(s)\ra_0^1 +\la s_x(s) s_y(s)\ra_0^2 \right],\label{eq:Cxy_renewal} 
\eea
where $\la s_x(s) s_y(s)\ra_0^{i}$ denotes the spatial correlation in the absence of resetting, starting from the configuration $\cal C_i, i=1,2.$ Unfortunately, it is hard to calculate the spatial correlations starting from the strongly inhomogeneous configurations $\cal C_1$ and $C_2$ 
and hence we cannot get any analytical expression for $C_{x,y}.$ However, we can get a qualitative idea about the same for the limiting case of large values of $r.$

For large $r \gg 1,$ the integral in Eq.~\eref{eq:Cxy_renewal}  is dominated by the contributions from small values of $s \ll r^{-1}.$ 
In particular, starting from $\cal C_1$ and $\cal C_2$, for sites $x,y$ away from the boundaries between the two halves, $s_x$ and $s_y$ evolves very slowly, and as a first approximation we can use the values at time $s=0$ to write,
\bea
C_{x,y} &\simeq & \frac 12 \left [\la s_x(0) s_y(0)\ra_0^1 +\la s_x(0) s_y(0)\ra_0^2 \right] \cr
&=&\left \{ \begin{array}{ll}
   \frac 12 & \textrm{if}\;  0 < x,y < \frac L2, \; \textrm{or} \; \frac L2 < x,y < L \cr
   0 & \textrm{otherwise.}
  \end{array} \label{eq:C_xy_0th}
  \right.
\eea
Figure \ref{fig:corr_cxy}(a) shows a plot of $C_{x,x+\Delta}$ as a function of $\Delta$ obtained from numerical simulations for different values of $x$ for a fixed (large) value of $r=5.$ Clearly, for $x>0,$ Eq.~\eref{eq:C_xy_0th} provides a reasonably well prediction. On the other hand, for $r \to 0,$ we expect that $C_{x,y} \to \frac 14,$ the equilibrium value, independent of $x,y.$ 

\begin{figure}[t]
 \centering
   \includegraphics[width=8.5 cm]{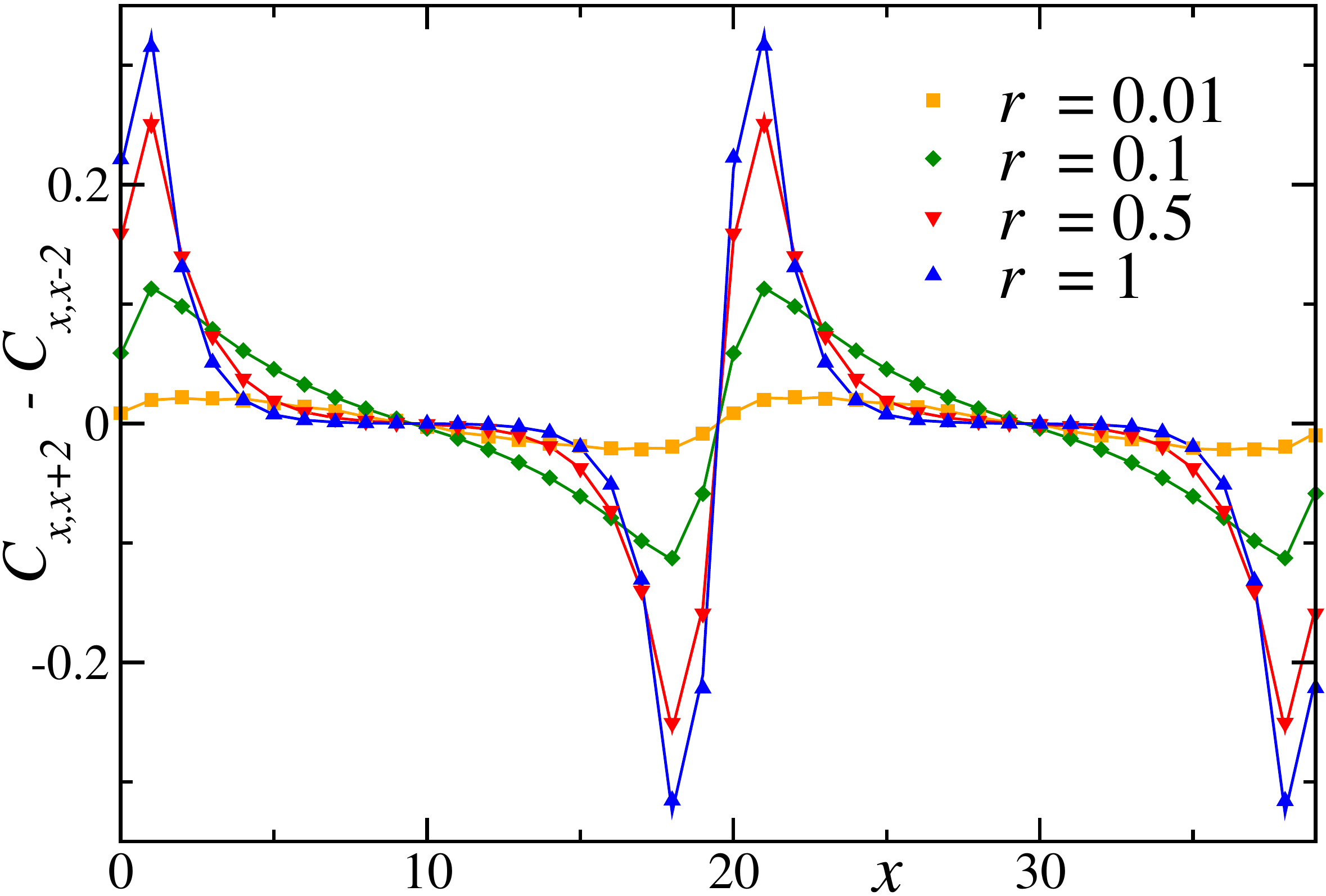}
 \caption{Spatial correlation in the stationary state: Comparison of $C_{x,x+2} - C_{x,x-2}$ (symbols) and $(r+2)[C_{x,x+1}-C_{x,x-1}] + \frac r2(\delta_{x,\frac L2 -1}-\delta_{x,0}+\delta_{x,L-1}-\delta_{x,\frac L2})$ (solid lines) as a function of $x$ for different values of $r.$ The data is obtained from numerical simulations using a lattice of size $L=40.$}
 \label{fig:corr_sisj}
\end{figure}

The non-trivial nature of the spatial correlation for finite values of $r$ is shown in Fig.~\ref{fig:corr_cxy}(b) where $C_{x,x+\Delta},$ measured from numerical simulations, is plotted for different values of $r.$
As $r$ decreases the correlation between the two halves increases.
To gain more information about the behaviour of $C_{xy}$ for intermediate values of $r$ we adopt a different approach.  From the master equation \eref{eq:ME_PC}, multiplying both sides of Eq.~\eref{eq:ME_PC} by $s_x s_y$ and summing over all configurations $\cal C$ we get the time-evolution
equation for $C_{x,y}.$ For $r_1=r_2=\frac r2,$ this equation reads,
\bea
\fl \frac{\id}{\id t} C_{x,y} = C_{x,y+1}+C_{x,y-1}+C_{x+1,y}+C_{x-1,y}-(r+4) C_{x,y} +\frac r2 \sum_{\cal C} s_x s_y (\delta_{\cal C,\cal C_1} + \delta_{\cal C,\cal C_2}) \cr
+\delta_{x+1,y} (2 C_{x,y}-C_{x,y-1}-C_{x+1,y})  +\delta_{x-1,y}(2 C_{x,y}-C_{x,y+1}-C_{x-1,y}). \label{eq:ME_Spatial_correl}
\eea
In the stationary state the left hand side vanishes, yielding a relation between $C_{xy}$ at different spatial points,
\bea
\fl C_{x,y+1}+C_{x,y-1}+C_{x+1,y}+C_{x-1,y}= (r+4) C_{x,y} -\frac r2 \sum_{\cal C} s_x s_y (\delta_{\cal C,\cal C_1} + \delta_{\cal C,\cal C_2})\cr
- \delta_{x+1,y} (2 C_{x,y}-C_{x,y-1}-C_{x+1,y})-\delta_{x-1,y}(2 C_{x,y}-C_{x,y+1}-C_{x-1,y}). \label{eq:Cxy_stationary}
\eea
While still not solvable exactly, this equation provides a simple relation between nearest neighbour and next nearest neighbour correlations. Substituting $y=x+1$ and $y=x-1$ in Eq.~\eref{eq:Cxy_stationary}, we have, respectively,
\bea
\fl \qquad C_{x,x+2}+C_{x-1,x+1} = (2+r) C_{x,x+1} - \frac r2 \sum_{\cal C} s_x s_{x+1} (\delta_{\cal C,\cal C_1} +\delta_{\cal C,\cal C_2} ), \label{eq:dKxx1dt} \\
\fl \qquad C_{x,x-2}+C_{x-1,x+1}= (2+r) C_{x,x-1}-\frac r2 \sum_{\cal C}  s_x s_{x-1} (\delta_{\cal C,\cal C_1} + \delta_{\cal C,\cal C_2}) \label{eq:dKxx-1dt}
\eea
for any value of $x.$ Subtracting Eq.~\eref{eq:dKxx-1dt} from Eq.~\eref{eq:dKxx1dt}, we get,
\bea 
\fl C_{x,x+2}-C_{x,x-2}&=&(2+r)(C_{x,x+1}-C_{x,x-1}) -\frac r2 \sum_{\cal C}  s_x (s_{x+1}-s_{x-1}) (\delta_{\cal C,\cal C_1}  + \delta_{\cal C,\cal C_2}) \cr
&=&(2+r)[C_{x,x+1}-C_{x,x-1}]+\frac{r}2 (\delta_{\frac{L}{2}-1,x}-\delta_{0,x}+\delta_{L-1,x}-\delta_{\frac{L}{2},x}), \label{eq:Cx_dif}
\eea
where, in the last step, we have used the fact that  the terms containing the $\delta_{\cal C,\cal C_1}$ and $\delta_{\cal C,\cal C_2}$ are non-zero only for four lattice sites, namely, $x=0,\frac{L}{2}-1,\frac{L}{2},$ and $x=L-1$. 
This relation provides a way of directly demonstrating the non-trivial correlation induced by the presence of resetting. For Ordinary SEP, the quantity ($C_{x,x+2}-C_{x,x-2}$) vanishes in stationary state for all values of $x,$ even for a finite lattice [see Eq.~\eref{eq:cxy_eq}]. On the other hand, Eq.~\eref{eq:Cx_dif} predicts a non-trivial $r-$dependent value for the ZCS.

We use numerical simulations to illustrate this non-trivial spatial correlation. Figure \ref{fig:corr_sisj} shows a plot of $C_{x,x+2}-C_{x,x-2},$ measured directly (symbols), along with the right hand side  of Eq.~\eref{eq:Cx_dif} (solid lines) as a function of $x.$  The curves become more and more inhomogeneous, particularly near the boundaries $x=0$ and $x=L/2$ between the two halves of the lattice, as the resetting rate $r$ is increased indicating strong spatial correlation in the system.  It is also consistent with the directly measured $C_{x,y}$ [see Fig.~\ref{fig:corr_cxy}(a)] which shows a big jump near the boundaries $x=0,L/2$ and hence resulting in a significant change in the difference also. Note that the difference also increases with $r$, in agreement with  Eq.~\eref{eq:Cx_dif}. 

%

\subsection{Temporal correlations} \label{subsec:time_correl}

The presence of stochastic resetting is expected to affect the temporal correlations of the system as it introduces additional time-scales. In this section we investigate the behaviour of the density auto-correlation in the ZCS. In particular, we focus on the two-point auto-correlation at site $x,$  
\bea
\fl \quad \cal G(t,t+\tau) \equiv \la s_x(t) s_x(t+\tau) \ra = \sum_{\cal C ,\cal C'} s_x(t+\tau) s_x(t)~ \cal P(\cal C', t+\tau| \cal C,t)~\cal P(\cal C, t| \cal C_0,0). \label{eq:G-def}
\eea
In the stationary state, $\cal G(t,t+\tau)$ is expected to depend only on $\tau.$ 

In the absence of resetting, the stationary (\ie, equilibrium) correlation $\cal G_0(\tau)$ can be calculated explicitly (see~\ref{sec:app}) and for $\rho=\frac 12$ it turns out to be,
\bea
\cal G_0(\tau) = \frac 14 [e^{-2 \tau} I_0(2 \tau)+ 1].
\eea
In the limit $\tau \to \infty,$  $s_x(t)$ and $s_x(t+\tau)$ decorrelate and the auto-correlation saturates to $\rho^2 = \frac 14.$ The approach to this value can be obtained from the above equation, and turns out to be algebraic in nature \cite{Spohn-book},     
\bea
\tilde \cal G_0(\tau) \equiv \cal G_0(\tau) - \frac 14 = \frac 1{8 \sqrt{\pi \tau}} +\cal O(\tau^{-1}). \label{eq:temp_correl_lt}
\eea 

\begin{figure}[t]
 \centering
 \includegraphics[width=8 cm]{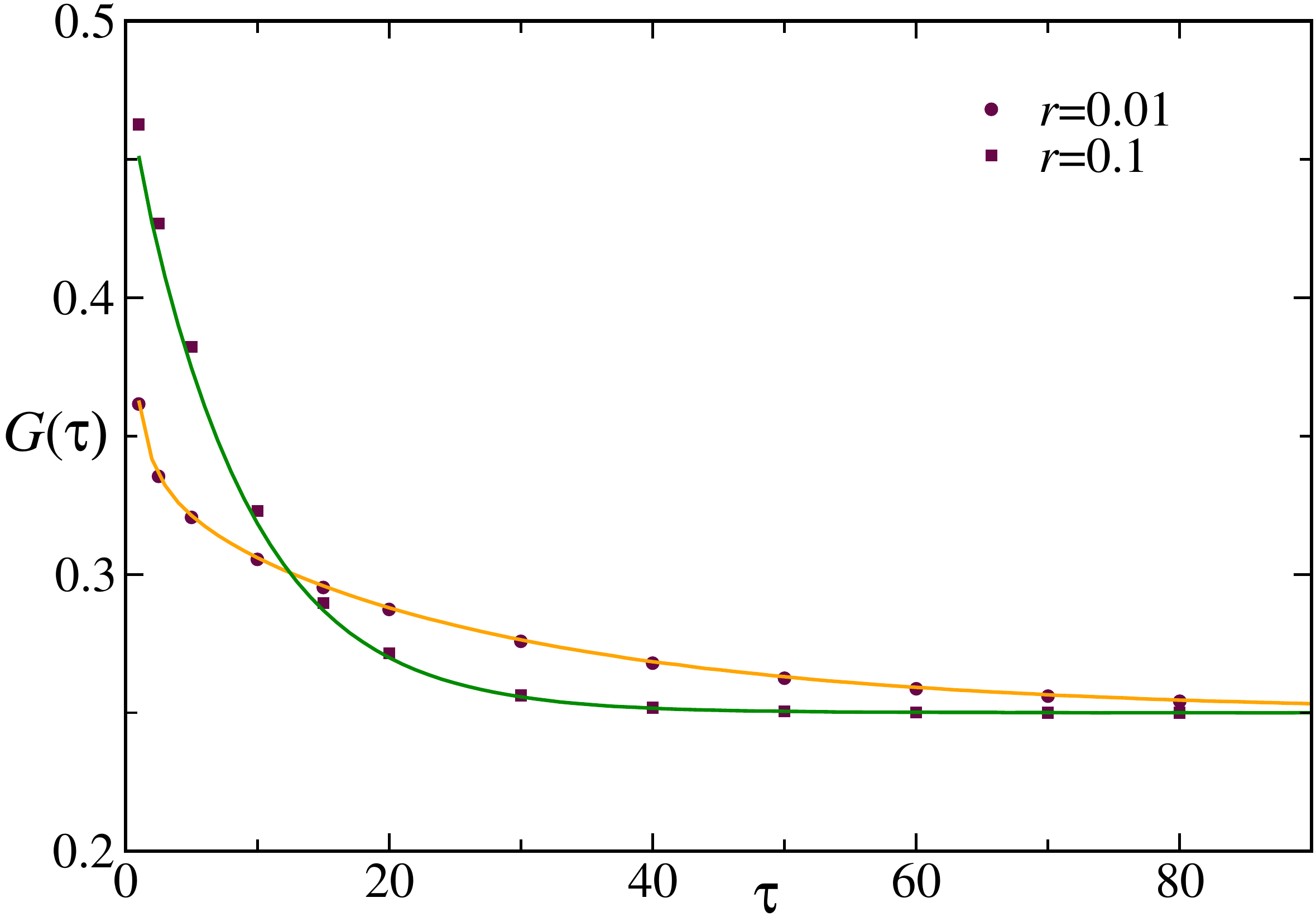}
 \caption{Numerical verification of Eq.~\eref{eq:G_tau_lt} for two different values of $r.$ The solid line represents left hand side of Eq.~\eref{eq:G_tau_lt}, while the dots represent the right hand side of the same equation.}
 \label{fig:Gtau_cmp}
\end{figure}

To understand the effect of the resetting on the density auto-correlation, let us first recall that, for any $t > t',$ the corresponding conditional probability $\cal P(\cal C', t'| \cal C,t)$ satisfies a renewal equation (see Eq.~\eref{eq:PC_sol2}),
\bea
\fl \cal P(\cal C, t| \cal C',t') = e^{-r(t-t')} \cal P_0(\cal C, t| \cal C',t') + \frac r2 \int_0^{t-t'}\id s~ e^{-r s}[\cal P_0(\cal C,s|\cal C_1,0 )+\cal P_0(\cal C,s|\cal C_2,0 )].\label{eq:PC_cond}
\eea
Note that here we have restricted to the case $r_1=r_2=\frac r2.$ 
Using Eq.~\eref{eq:PC_cond} in Eq.~\eref{eq:G-def} and performing the sums over the configurations we get,
\bea
\fl \qquad \cal G(t,t+\tau) &=& e^{-r(t+\tau)} \cal G_0(t,t+\tau) + \frac r2 e^{-r \tau} \int_0^t \id s ~e^{- r s} [\cal G^1_0(s,s+\tau) +\cal G^2_0(s,s+\tau) ] \cr
&+& \frac r2 e^{-rt} \rho_0(x,t) \int_0^\tau \id s~ e^{-r s} [\rho^1_0(x,s) +\rho^2_0(x,s)] \cr
&+&  \frac{r^2}4 \int_0^\tau \id s~ e^{-r s} [\rho^1_0(x,s) +\rho^2_0(x,s)] \int_0^t \id s'~ e^{-r s'} [\rho^1_0(x,s') +\rho^2_0(x,s')], \label{eq:Gt_2}
\eea
where $\cal G^i_0(t,t') = \la s_x(t) s_x(t') \ra_0$ denotes the auto-correlation in the absence of resetting starting from configuration $\cal C_i$ and  $\cal G_0(t,t')$ denotes the same starting from $\cal C_1$ and $\cal C_2$ with equal probability (which is our chosen initial condition). Similarly, $\rho_0(x,t)$ denotes the average density at site $x$ at time $t$, starting from this chosen initial condition while $\rho_0^i(x,t)= \sum_{\cal C} s_x \cal P_0(\cal C,s|\cal C_i,0)$ denotes the density starting from configuration $\cal C_i,$ in the absence of resetting. Now, using the results of Sec.~\ref{sec:density}, it is easy to see that, $\rho^1_0(x,t)+ \rho^2_0(x,t) =1$ at any time $t.$ Moreover, for our choice of initial condition ($\cal C_1$ and $\cal C_2$ with equal probability) $\rho_0(x,t)=1/2.$ Using these in Eq.~\eref{eq:Gt_2} we get a renewal equation for $\cal G(t,t+\tau),$ 
\bea
\fl \cal G(t, t+\tau) = e^{-r(t+\tau)} \cal G_0^{0}(t,t+\tau) + \frac{r}2 e^{-r \tau} \int_0^t \id s~ e^{-r s}\bigg[\cal G_0^1(s,s+\tau)+\cal G_0^2(s,s+\tau)\bigg]\cr
+ \frac 14 (1-e^{-r \tau}). \label{eq:G_tau}
\eea
In particular, in the stationary state $t \to \infty,$ the auto-correlation depends on $\tau$ only,
\bea
\fl \qquad \, \cal G(\tau) = \frac{r}2 e^{-r \tau} \int_0^\infty \id s~ e^{-r s}\bigg[\cal G_0^1(s,s+\tau)+\cal G_0^2(s,s+\tau)\bigg]+ \frac 14 (1-e^{-r \tau}). \label{eq:G_tau_lt}
\eea
To verify this relation, using numerical simulations we measure $\cal G_0^1(s,s+\tau)$ and $\cal G_0^2(s,s+\tau)$ as a function of $s,$ for different values of $\tau,$ in the absence of resetting. Then performing the integration over $s$ numerically, we can compute the right hand side of Eq.~\eref{eq:G_tau_lt}. The data obtained thus are shown as the symbols in Fig.~\eref{fig:Gtau_cmp} for two different values of $r.$ On the other hand, we also measure the stationary correlation $\cal G(\tau)$ in the presence of resetting directly which are shown as the black curves in the same figure. Clearly, the two measurements agree perfectly which verifies Eq.~\eref{eq:G_tau_lt}.

\begin{figure}[t]
 \centering
 \includegraphics[width=15 cm]{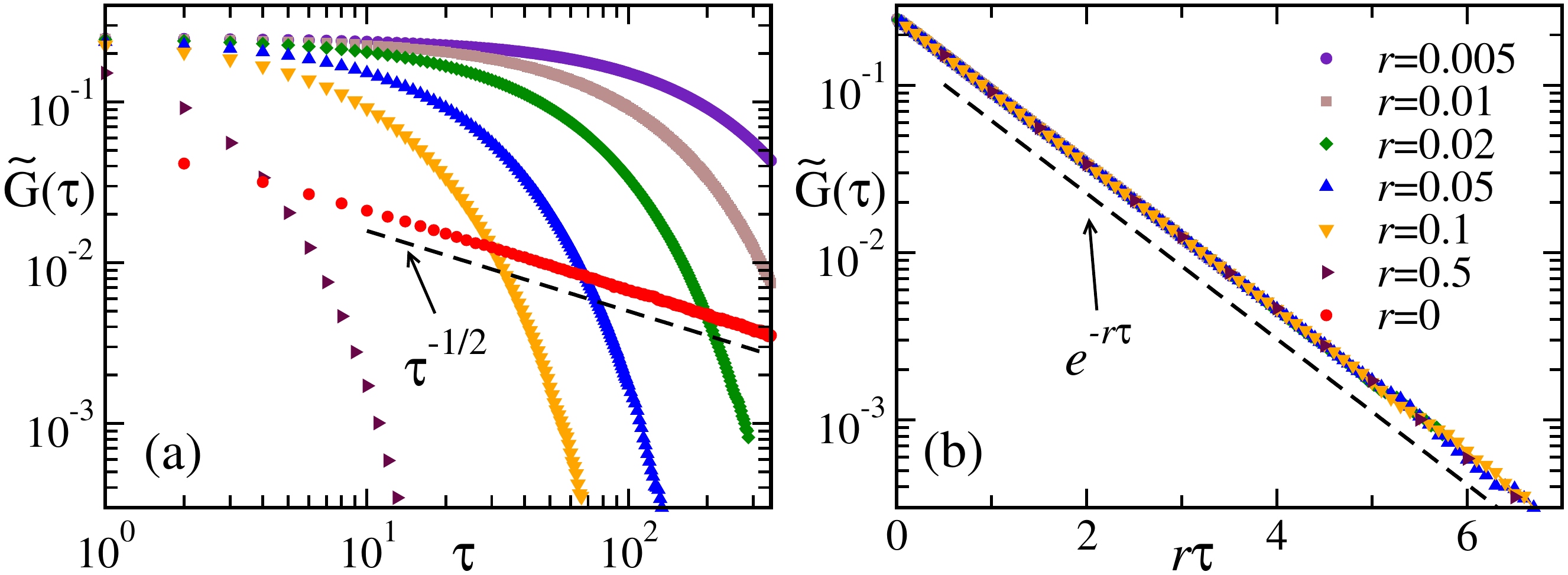}
 \caption{ Figure (a) shows the graph for different values of $r$. We can see the crossover with the $r=0$ case. (b) shows the same plot but as a function of scaled variable $r \tau$.}	 
 \label{fig:corrt}
\end{figure}

We can make further progress for small values of $r.$ In this case the integral in Eq.~\eref{eq:G_tau_lt} is dominated by large values of $s.$ In this regime $\cal G^i_0(s,s+\tau)$ is expected to become independent of $s;$ as a first approximation we can use the stationary correlation $\cal G_0(\tau)$ given by Eq.~\eref{eq:temp_correl_lt}. Then we get, from Eq.~\eref{eq:G_tau_lt},
\bea
\cal G(\tau) \simeq e^{-r \tau} \left[\cal G_0(\tau) - \frac 14 \right] + \frac 14 = \frac 14 \left[e^{-(r+2)\tau} I_0(2\tau)+1 \right]. \label{eq:Gtau_stat}
\eea
Using the asymptotic behaviour of the Bessel function it is easy to see that the connected correlation $\tilde \cal G(\tau)$ is expected to decay exponentially for large $\tau.$ This is shown in Fig.  \ref{fig:corrt}(a) where $\tilde \cal  G(\tau)$ for different values of $r,$ obtained from numerical simulations, are plotted as a function of $\tau.$ For easy reference we have also included the data for $r=0$ which shows the $\tau^{-1/2}$ decay. Figure \ref{fig:corrt}(b) shows the data for non-zero $r$  plotted as a function of $r \tau$ which show a perfect collapse according to the predicted exponential decay. 

This exponential temporal correlation is a typical feature of systems with stochastic resetting and originates from the fact that in the presence of resetting, the configurations are correlated only when they occur between two consecutive resetting epochs. Similar results have been observed in the context of single particle resetting with one rate in Ref.~\cite{Oshanin2018}. 
For our present case of SEP with dichotomous resetting we see that while the resetting creates strong spatial correlations, it effectively erases the non-trivial temporal correlation present in ordinary SEP, replacing it with a faster, exponential decay which depends only on the resetting rate $r.$

\subsection{Response to perturbation} \label{subsec:perturb}

How a system responds to an external perturbation plays an important role in characterizing the properties of the system. For a small perturbation around equilibrium, the response is predicted by the famous fluctuation-dissipation theorem (FDT) which gives an explicit form for the susceptibility in terms of equilibrium correlations \cite{Kubo}. For systems away from equilibrium,  the linear response formula is different, notably, with the addition of a `frenetic' contribution \cite{Maes2009}.

To further characterize the zero-current state for $r_1=r_2,$ and illustrate its nonequilibrium nature we investigate the response of the system to an external perturbation, and compare it with the same for equilibrium SEP. In the following we consider one of  the simplest possible perturbations, namely, a biasing field $\ve$ across the central bond. This is implemented by a change in the hop rates {\it across the central bond},
\bea
10 \mathop{\rightleftharpoons}^{p}_{q} 01,
\eea
where $p/q = e^{\ve},$ consistent with local detailed balance \cite{kls}. The hop rate for all the other bonds remains unchanged. 


This bias  generates a non-zero diffusive current, for both equilibrium SEP and the zero-current nonequilibrium state. For a small biasing field, the average current generated  $\la \Jd(t) \ra_\ve$  is expected to be proportional to the field $\ve$ in both the cases with a susceptibility $\chi= \la \Jd(t) \ra_\ve/ \ve$ which depends on the fluctuations in the unperturbed state. In the equilibrium case, the response is given by the Kubo formula, which predicts that the susceptibility is proportional to the variance of the current in absence of the perturbation, 
\bea
\chi_{\textrm{eq}}= \frac 12 \la \Jd^2 \ra_0 \equiv \la J_0^2 \ra. \label{eq:chi_eq}
\eea
where $\la J_0^2 \ra$ is the variance in the equilibrium state, \ie, both in the absence of the resetting and the perturbation, given by Eq.~\eref{eq:J02_J04}. From the above equation and Eq.~\eref{eq:J02_J04}, it is clear that the average current generated due to the perturbation to ordinary SEP grows as $\sqrt t$ with time. 

To predict the response of the ZCS to the perturbation one has to take recourse to the nonequilibrium response theory \cite{Maes2009}. Using path integral formalism, the linear response can be expressed in terms of correlations in absence of the perturbation,
\bea
\chi_{\textrm{r}} = \frac 1 2 \la \cal S(\omega) \Jd \ra - \la \cal D(\omega)  \Jd \ra. \label{eq:chi_r}
\eea
Here $\cal S(\omega)$ and $\cal D(\omega)$ respectively denote the {\it excess} entropy and  dynamical activity generated by the perturbation along the trajectory $\omega$ during the interval $[0,t];$ the correlations are calculated by averaging over all possible trajectories in the zero-current state, \ie, in the presence of the resetting but in the absence of the perturbations. The path dependent quantities  $\cal S$ and $\cal D$ are the time-antisymmetric and time-symmetric parts of the path-action calculated with respect to the unperturbed path weights and can be explicitly computed for any stochastic process given the perturbation protocol. For detailed prescription of how to compute these quantities see Refs.~\cite{Maes2009,update_non_eq}.
Here we just give the expressions for the particular perturbation under consideration. The excess entropy is independent of the specific form of $p$ and $q,$ and is given by,
\bea
\cal S(\omega) &=& \Jd .\label{eq:S}
\eea
Hence, the first term in Eq.~\eref{eq:chi_r} is nothing but the Kubo-term, measuring the variance in the unperturbed state. On the other hand, the excess dynamical activity depends explicitly on $p$ and $q$ 
\bea
\cal D(\omega) &=& - \frac 12 \cal N \frac \id{\id \ve}(\ln pq)\Bigg|_{\ve=0} +  \frac {\id p}{\id \ve}\Bigg|_{\ve=0} t_{10} + \frac {\id q}{\id \ve}\Bigg|_{\ve=0} t_{01} .\label{eq:D}
\eea
Here $\cal N$ denotes the {\it total} number of jumps across the central bond (both towards right and left). Moreover, $t_{01}$ (respectively, $t_{10}$) denotes the total time in $[0,t]$ during which the local configuration is $01$ (respectively, $10$) in the central bond, \ie, $s_{\frac L2-1}=0,s_{\frac L2}=1$ (respectively, $s_{\frac L2-1}=1,s_{\frac L2}=0$). 

\begin{figure}[t]
 \centering
 \includegraphics[width=7.4 cm]{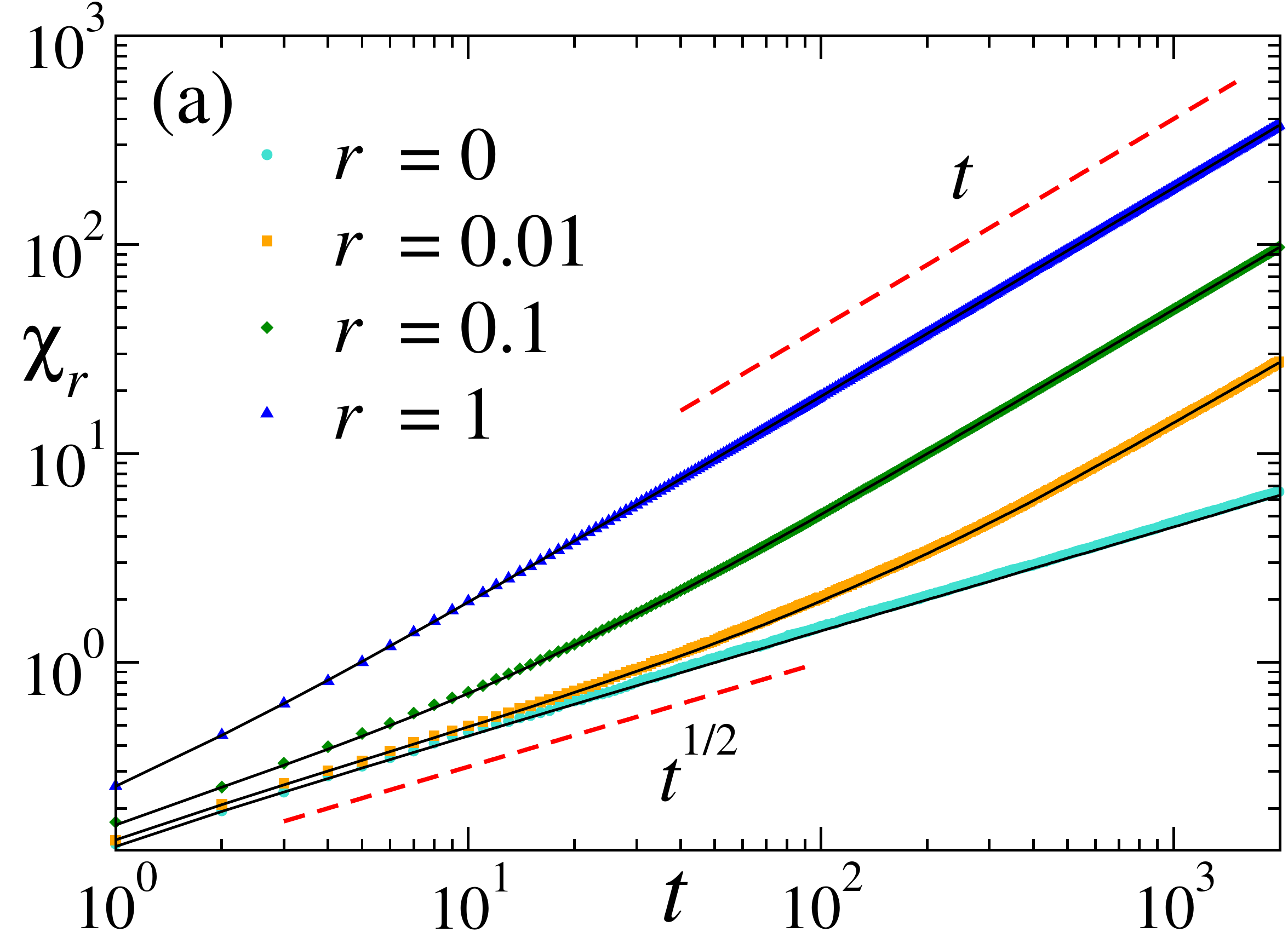}  \includegraphics[width=7.6 cm]{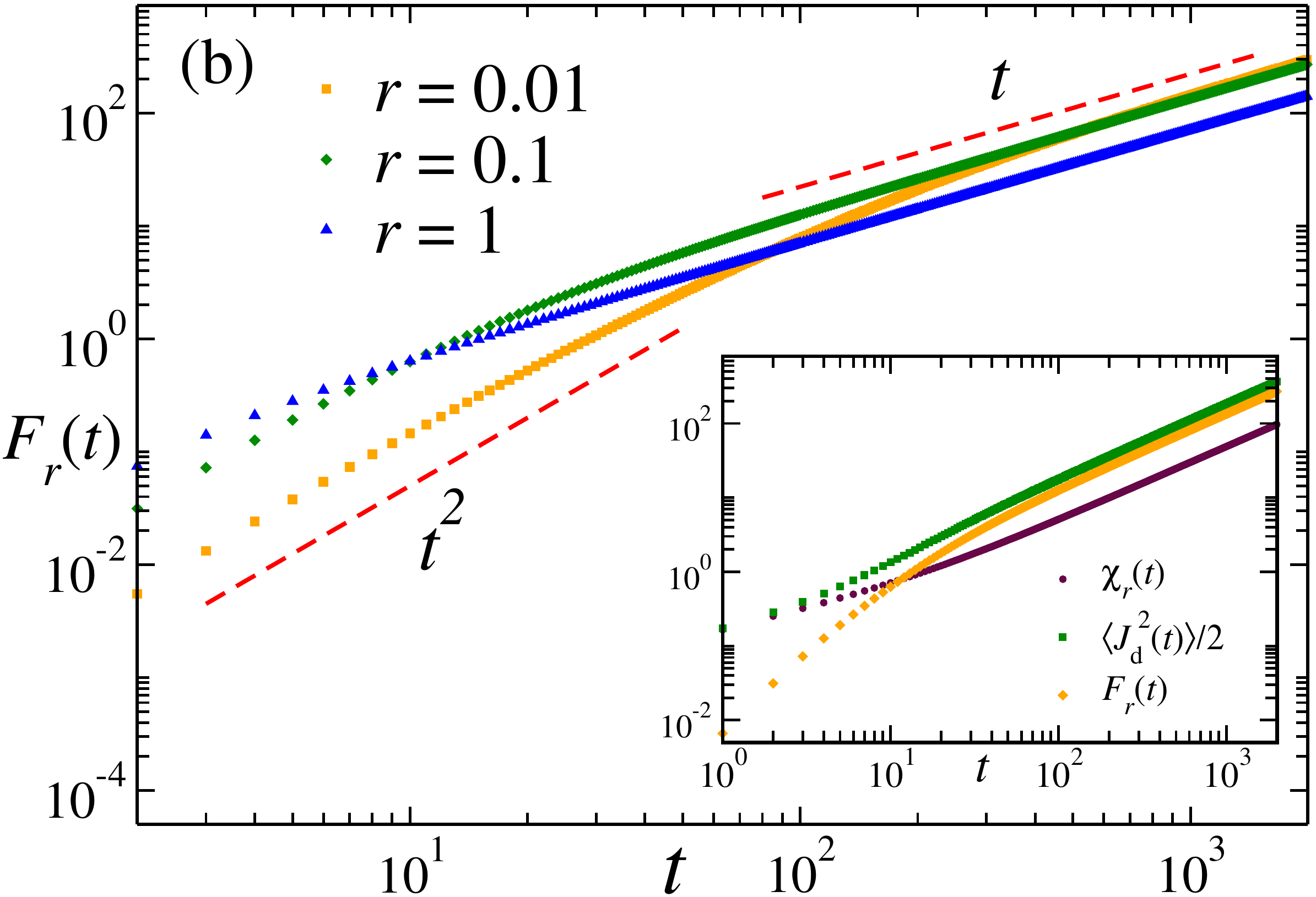}  
\caption{Linear response in ZCS: (a) Plot of  the nonequilibrium susceptibility $\chi_r$ $vs$ $t$ for different values of $r.$ The symbols correspond to the predicted response Eq.~\eref{eq:noneq_sus} measured in the unperturbed system, while the solid lines corresponds to direct measurement of susceptibility in the presence of a perturbation of strength $\ve=0.1$. For $r>0,$ $\chi_r$ grows $\sim t$ at large times while for $r=0$ it grows as $\sqrt{t}.$ 
(b) Plot of the frenetic component $F_r(t)= \frac 12 (\la t_{10} \Jd \ra - \la t_{01} \Jd \ra)$  which shows a crossover from a $t^2$ behaviour at short-times to a linear growth at late times. The inset shows the entropic and frenetic components along with the total $\chi_r$ for a fixed $r=0.1.$ All the simulations are done on a lattice of size $L=1000.$}
 \label{fig:resp}
\end{figure}

To proceed further, we need to specify $p$ and $q.$ For the sake of simplicity, we consider $p = e^{\ve/2}$ and $q=e^{-\ve/2}.$ 
Note that the this formula depends crucially on the choice of $p,q$ but for our purpose it suffices to consider this simple choice. In this case $pq=1$ and  Eq.~\eref{eq:D} simplifies leading to the nonequilibrium susceptibility,
\bea
\chi_r (t) = \frac 12 [ \la \Jd^2(t) \ra -  \la t_{10} \Jd \ra + \la t_{01} \Jd \ra]. \label{eq:noneq_sus}
\eea

The first term is nothing but the variance of the diffusive current in the presence of the dichotomous resetting which we have already calculated in Sec.~\ref{sec:current} [see Eq.~\eref{eq:Jd2} with $r_1=r_2$]. The frenetic component $F_r(t)= \frac 12 (\la t_{10} \Jd \ra - \la t_{01} \Jd \ra)$ involves correlation of the current with time-symmetric quantities $t_{01}$ and $t_{10}$ and we take recourse to numerical simulations to measure these. We also measure the susceptibility  directly by applying the perturbation and calculating $\chi_r = \la \Jd \ra_\ve/\ve$. Figure~\ref{fig:resp}(a) compares the  $\chi_r$  predicted by Eq.~\eref{eq:noneq_sus} (symbols) with the directly measured $\chi_r $ (solid lines) for different values of $r.$ For $r>0,$ at late times, the susceptibility grows linearly with $t$ which is drastically different than the $r=0$ scenario, where the susceptibility $\sim \sqrt{t}.$ Note that, at short-times $t \ll r^{-1}$, the system shows an equilibrium-like behaviour with $\chi_r \sim \sqrt{t}.$

It is interesting to look at the frenetic component $F_r(t)$ separately. Figure \ref{fig:resp}(b) shows plots of $F_r(t)$ for different values of $r.$ At short-times $t \ll r^{-1}$ it shows a $t^2$ behaviour, and is thus negligible compared to the entropic component which $\sim \sqrt{t}$ in this regime [see Eq.~\eref{eq:Jd2} with $r_1=r_2$]. Hence the total response in this regime is dominated by the entropic component only, which is expected, as at short-times the effect of resetting is not visible, and the system remains close to equilibrium. At late-times, $F_r(t)$ shows a $\sim t$ behaviour, similar to the entropic part and the total response also becomes linear. The inset in Fig.~\ref{fig:resp}(b) shows the entropic and frenetic components along with the $\chi_r$ for a fixed $r$; clearly, the negative contribution from the frenetic component reduces the slope of the total response.

%
%

To summarize, the study of the linear response in the ZCS shows two important features which distinguish it from equilibrium SEP. First, the same perturbation, namely, a small driving field gives rise to very different currents in the two cases -- in presence of the resetting the current grows much faster. Secondly, the linear response has a frenetic contribution which competes with the traditional Kubo term and reduces the slope of the susceptibility (as a function of time).

\section{Conclusions}\label{sec:conclusions}  

In this article we have studied the behaviour of symmetric exclusion process in the presence of dichotomous stochastic resetting. The dichotomous resetting is implemented by resetting the  system to either of the two specific configurations where all the particles are in the left (respectively, right) half of the system, with rates $r_1$ (respectively $r_2$). The presence of the dichotomous resetting leads to intriguing dynamical and stationary properties, which we have characterized. We have exactly calculated time-evolution of the density profile which remains inhomogeneous in the stationary state. 

The primary quantity of interest is the diffusive particle current across the central bond. We show that for any $r_1 \ne r_2$ the current grows linearly with time, with a coefficient proportional to $(r_1-r_2).$ In the long-time limit the distribution of the current converges to a Gaussian distribution, while at the short-time regime there are strong non-Gaussian fluctuations which we characterize via skewness and kurtosis. We demonstrate that both skewness and kurtosis approach the Gaussian limit with an algebraic decay in time.

A special scenario arises when $r_1=r_2.$ In this case, the system reaches a stationary state where the density profile is uniform and there is no particle current flowing through the system. Nevertheless, the system stays far away from equilibrium and we characterize this zero current nonequilibrium state by computing the spatial and temporal density correlations. We show that while the presence of the resetting induces non-trivial spatial correlations, it also reduces the temporal correlations -  instead of an algebraic temporal decay of the auto-correlation (which is seen for equilibrium SEP), we get an exponential decay in the presence of the resetting. Finally, we also study the response of the ZCS to a small perturbation and show that the response is linear in time instead of the $\sqrt{t}$ behaviour expected in equilibrium.

We conclude with the final remark that the simple dichotomous resetting protocol leads to a more rich and complex behaviour compared to resetting with a single rate and it would be interesting to study the effect of such resetting protocols for single particle systems as well as more complicated interacting systems.

\ack
The authors thank Riddhipratim Basu, Aanjaneya Kumar and Christian Maes for helpful discussions. O. S. acknowledges the Inspire grant from DST, India and support from Raman Research Institute where he worked as a visiting student and where this work was carried out.  U. B. acknowledges support from Science and Engineering Research Board, India under Ramanujan Fellowship (Grant No. SB/S2/RJN-077/2018).

\appendix

\section{Density auto-correlation for ordinary SEP}\label{sec:app}

In this Section we briefly revisit the temporal auto-correlation of equilibrium SEP, \ie, in the absence of resetting. In particular, we look at the density auto-correlation at site $x,$ 
\bea
\cal G^{0}(t) = \la s_x(0) s_x(t) \ra = \la s_x(t)| s_x=1,t=0 \ra .
\eea
Clearly, it is nothing but the density $\rho_x(t)$ at site $x,$ averaged over all possible initial configurations with $s_x=1.$ 
For SEP without resetting, we know that, starting from any initial density profile $\phi_0(y),$ the density at time $t$ is given by,
\bea
\rho_x(t) = \frac{1}{2} + \frac{1}{L}\left[ \displaystyle \sum_{n=1}^{L-1} \tilde\phi_0(n) \enskip e^{-\lambda_n t} \enskip e^{-i\frac{2{\pi}nx}{L}} \right], \label{eq:rho_x_t}
\eea 
where, $\tilde\phi_0(n)$ is the Fourier transform of $\phi_0(y).$ To compute the auto-correlation, we need the initial profile such that the density at site $x$ is 1. In equilibrium, for a large system size $L$, this is given by,
\bea
\phi_0(y)= \left \{  \begin{array}{cc}
                      1  \qquad \textrm{when} \qquad y=x \cr
                     \rho \qquad \textrm{when} \qquad y \neq x,
                    \end{array}
                    \right.
\eea
where, $\rho$ is the global particle density of the system. The corresponding Fourier transform is given by, 
\bea
\tilde\phi_0(n) &= \displaystyle \sum_{y=0}^{L-1} \enskip e^{i\frac{2{\pi}ny}{L}} \phi_0(y) = e^{i\frac{2{\pi}nx}{L}}+ \rho \displaystyle \sum_{y \neq x}^{L-1} \enskip e^{i\frac{2{\pi}ny}{L}} \cr
&= (1-\rho) e^{i\frac{2{\pi}nx}{L}}. 
\label{eq:phi_tilde_n}
\eea 
Subsituting Eq.~\eref{eq:phi_tilde_n} in Eq.~\eref{eq:rho_x_t}, we get for $\rho=\frac 12$,
\bea
\rho_x(t) 
&= \frac{1}{4}\left(1 + \frac 1{L} \displaystyle \sum_{n=1}^{L-1}  \enskip e^{-\lambda_n t}\right).
\eea
For large system size $L$, we can convert the summation to integral as done before in Eq.~\eref{eq:Jdav}. We get,
\bea
\cal G^{0}(t)&=\frac 14+ \frac{1}{8 \pi} \int_0^{2 \pi} dq \enskip e^{-\lambda_q t} \cr
&= \frac{1}{4}[1+ \enskip e^{-2t} I_0(2t)],
\eea
where $I_0$ is the modified Bessel function of the first kind. As expected, at large time $t \to \infty$, the correlation decays to $\rho^2= \frac 14.$ The approach to this stationary value can be obtained from the asymptotic behaviour of $I_0,$
\bea
\cal G^{0}(t) -\frac{1}{4} = \frac{1}{8 \sqrt{\pi t}}. 
\eea
which is quoted as Eq.~\eref{eq:temp_correl_lt} in Sec.~\eref{subsec:time_correl} of the main text.

\end{document}